\documentclass[twocolumn, final]{IEEEtran}

\usepackage{amsmath, amsfonts,amssymb, latexsym,epsfig,color,rotating,paralist,algorithmic,multirow,
times,float,subfigure,caption,algorithm,verbatim,framed,empheq}

\newtheorem{theorem}{Theorem}
\newtheorem{proposition}{Proposition}
\newtheorem{corollary}{Corollary}
\newtheorem{definition}{Definition}
\newtheorem{result}{Result}

\newcommand{\nn}{\nonumber}
\newcommand{\ole}{\stackrel{\triangle}{=}}

\newcommand{\uL}{\underline{L}}
\newcommand{\uA}{\underline{A}}
\newcommand{\nm}{L}
\renewcommand{\i}{\mathbf{i}}
\renewcommand{\j}{\mathbf{j}}

\newcommand{\gtptwo}{\underset{\text{\tiny TP2}}{\geq}}

\newcommand{\sv}{\sigma}
\newcommand{\iter}{l}
\newcommand{\Iter}{L}
\newcommand{\impdensity}{q}

\newcommand{\cons}{{\text{\bf Cons}}}

\newcommand{\lb}{\alpha}
\newcommand{\ub}{\beta}

\newcommand{\identity}{I}
\newcommand{\dvar}[2]{\|{#1}-{#2}\|_{\text{\tiny{TV}}}}

\renewcommand{\(}		{\left(}
\renewcommand{\)}		{\right)}

\newcommand{\F}{\mathcal{F}}
\newcommand{\dob}{\rho}
\newcommand{\bayes}{\mathcal{B}}
\newcommand{\filter}{T}
\newcommand{\obs}{y}

\newcommand{\state}{x}
\newcommand{\statespace}{\{1,2,\ldots,\statedim\}}
\newcommand{\obspace}{\{1,2,\ldots,\obsdim\}}
\newcommand{\statedim}{X}
\newcommand{\obsdim}{{Y}}
\newcommand{\one}{\mathbf{1}}

\newcommand{\statem}{F}

\newcommand{\map}{\state^{\text{MAP}}}
\newcommand{\lmap}{\underline{\state}^{\text{MAP}}}
\newcommand{\umap}{\bar{\state}^{\text{MAP}}}

\newcommand{\argmin}{\operatorname{argmin}}
\newcommand{\argmax}{\operatorname{argmax}}  

\newcommand{\levels}{g}

\newcommand{\oprob}{B}
\newcommand{\tp}{P}
\newcommand{\utp}{\bar{\tp}}
\newcommand{\ltp}{\underline{\tp}}

\newcommand{\diag}{\text{diag}}

\newcommand{\prob}{\mathbb{P}}
\newcommand{\E}                 {\mathbb{E}}
\newcommand{\reals}{\mathbb{R}}

\newcommand{\belief}{\pi}
\newcommand{\hbelief}{\hat{\belief}}
\newcommand{\hlbelief}{\hat{\lbelief}}

\newcommand{\albelief}{\underline{\gamma}}

\newcommand{\hlabelief}{\hat{\albelief}}

\newcommand{\Belief}{\Pi}
\newcommand{\ubelief}{\bar{\pi}}
\newcommand{\lbelief}{\underline{\pi}}
\newcommand{\tbelief}{\tilde{\belief}}

\newcommand{\pdf}{P}
\newcommand{\p}{\prime}
\newcommand{\filterd}{\sigma}

\newcommand{\onoise}{v}
\newcommand{\onoisevar}{\sigma^2_\onoise}

\newcommand{\alg}{\mathcal{A}}

\newcommand{\gr}{\geq_r}
\newcommand{\lr}{\leq_r}

\newcommand{\lR}{\preceq}
\newcommand{\gR}{\succeq}

\newcommand{\gs}{\geq_s}
\newcommand{\ls}{\leq_s}
\newcommand{\lowdim}{R}

\newcommand{\lmean}{\underline{\state}}
\newcommand{\mean}{{\hat{\state}}}
\newcommand{\umean}{\bar{\state}}

\newcommand{\mat}{M}

\newcommand{\beq}{\begin{equation}}
\newcommand{\eeq}{\end{equation}}

\begin{document}
\title{
Reduced Complexity Filtering with Stochastic Dominance Bounds:
 A    Convex Optimization Approach} 

\author{Vikram Krishnamurthy,  {\em Fellow, IEEE}  \and Cristian R. Rojas, {\em Member, IEEE}
\thanks{Vikram Krishnamurthy is
 with the Department of Electrical and Computer
Engineering, University of British Columbia, Vancouver, V6T 1Z4, Canada. 
(email:  vikramk@ece.ubc.ca). This research was partially supported by NSERC, Canada.
Cristian R. Rojas is with the ACCESS Linnaeus Centre and Automatic Control Lab, KTH Royal Institute of Technology, SE 100 44 Stockholm, Sweden. (email: crro@kth.se).
}}

\maketitle

\begin{abstract}
This paper uses stochastic dominance principles to construct upper and lower sample path bounds for Hidden Markov
Model (HMM)  filters.  Given a HMM, by using convex optimization methods for nuclear norm minimization
with copositive constraints,  we construct low rank stochastic  matrices $\ltp$ and $\utp$ 
 so that the optimal filters using $\ltp,\utp$  provably lower  and upper bound 
(with respect to a partially ordered set)
the true filtered
distribution at each time instant. Since $\ltp$ and $\utp$ are low rank (say $\lowdim$), the computational cost of evaluating the filtering
bounds is $O(\statedim \lowdim)$ instead of $O(\statedim^2)$. A Monte-Carlo importance sampling filter is presented that exploits
these upper and lower bounds to estimate the optimal posterior.  Finally, using the Dobrushin coefficient, explicit bounds are given
on the variational norm between the true posterior and the upper and lower bounds.
\end{abstract}

\section{Introduction}\label{sec:intro}
This paper is motivated by the 
  filtering problem involving estimating a large dimensional  finite state Markov chain given noisy observations. With $k$ denoting discrete time,
consider an $\statedim$-state discrete time Markov chain $\{\state_k\}$ observed via a noisy process $\{\obs_k\}$.
Here $\state_k \in \statespace$  where $\statedim$ denotes the dimension of the state space.  
Let  $\tp$ denote the $\statedim \times \statedim$ transition matrix and $\oprob_{\state\obs }= \prob(\obs_k = \obs|\state_k=\state)$ denote the observation likelihood probabilities. With $\obs_{1:k}$ denoting  the sequence of observations from time 1 to $k$,    define the posterior state probability mass function 
\begin{align}
 \belief_k(i) &= \pdf(\state_k=i|\obs_{1:k}), \; i \in \statespace, \nn \\
  \belief_k &= \begin{bmatrix}  \belief_k(1), \ldots, \belief_k(\statedim) \end{bmatrix}^\p. 
\label{eq:beliefvector}
\end{align}
%
 It is well known \cite{CMR05,EAM95} that the optimal Bayesian filter (Hidden Markov Model filter) for computing the $\statedim$-dimensional posterior vector $\belief_k$ at each time $k$ is of the form 
 \beq \label{eq:filter}
   \belief_{k+1} =  \filter(\belief_k,\obs_{k+1};\tp) \ole \frac{\oprob_{\obs_{k+1}} \tp^\p \belief_k} { \one^\p  \oprob_{\obs_{k+1}} \tp^\p \belief_k} 
   .\eeq
%
  Here, $ \oprob_{\obs} = \diag(\oprob_{1\obs},\ldots,\oprob_{\statedim y})$ is a diagonal $\statedim$-dimensional matrix of observation likelihoods and $\one$ denotes the $\statedim$-dimensional column vector of ones.

Due to the  matrix-vector multiplication $\tp^\p \belief_{k}$ in (\ref{eq:filter}),
the
computational cost for evaluating the posterior $\belief_{k+1}$  at each time $k$ is  $O(\statedim^2)$.
This  quadratic computational cost $O(\statedim^2)$ can be excessive for large state space dimension~$\statedim$. 

\subsection*{Motivation and Main Results}
This paper addresses the question:  {\em Can the optimal filter be approximated with reduced complexity filters
with provable sample path bounds?}
We derive reduced-complexity filters with computational cost  $O(\lowdim\statedim)$ where $\lowdim \ll \statedim$.
There are four main results in this paper.
\\
1. {\em Stochastic Dominance Bounds}:  
Theorem \ref{thm:main} presented in Sec.\ref{sec:copositive} asserts that for any  transition matrix $\tp$, one can construct two new transition
matrices $\ltp$ and $\utp$, such that
$\ltp \lR \tp \lR \utp$.
 Here $\lR$ denotes a {\em copositive ordering} defined in Section~\ref{sec:copositive}.
The Bayesian filters using  $\ltp$ and $\utp$, are guaranteed to  sandwich 
 the true posterior distribution $\belief_k$ at any time  $k$ as 
\beq  \filter(\belief_{k-1},\obs_k;\ltp) \lr \filter(\belief_{k-1},\obs_k,\tp) \lr \filter(\belief_{k-1},\obs_k;\utp)    \label{eq:poset} \eeq
 where $\filter(\cdot)$ denotes the filtering recursion (\ref{eq:filter}) and $\lr$ denotes monotone likelihood ratio (MLR) stochastic dominance  defined in Section~\ref{sec:copositive}.
 What (\ref{eq:poset}) says  is that at any time $k$, the true posterior $\belief_k =  \filter(\belief_{k-1},\obs_k,\tp) $ can be sandwiched in the partially ordered set specified by
 the above stochastic dominance constraints.
 Moreover, if $\tp$ is a TP2 matrix\footnote{TP2 matrices are defined in Definition \ref{def:tp2}.},  this statement can be globalized to say that if $\lbelief_0 \lr \belief_0 \lr \ubelief_0$, then
\beq  \lbelief_k \lr \belief_k \lr \ubelief_k , \quad \text{ for all } k \label{eq:globalize}
\eeq  where
 $\lbelief_k$ and $\ubelief_k$ denote the posteriors computed using  $\ltp$ and $\utp$.

The MLR stochastic order $\lr$ used in (\ref{eq:poset}) and (\ref{eq:globalize})  is a partial order on the
 set of  distributions.
A crucial property  of  the MLR order  is that it is
 closed under conditional expectations. This makes it 
very useful in  Bayesian estimation \cite{Kri11,Lov87,Whi79}.
An important consequence of the sandwich result  (\ref{eq:globalize}) is that  the conditional mean state estimates are sandwiched as
$\lmean_k \leq \mean_k \leq  \umean_k$ for all time $k$. Indeed, the second and all higher moments also are sandwiched.

Finally,
in Sec.\ref{sec:multi} we generalize the above result to multivariate POMDPs by using the multivariate TP2 stochastic order. Such multivariate HMMs provide  a useful example of large scale HMMs.
The TP2 order was pioneered by Karlin \cite{KR80}, see also Whitt's classic paper \cite{Whi82}.
\\
2. {\em Construction of low rank transition matrices via nuclear norm minimization}: Sec.\ref{sec:convex} uses  state-of-the-art convex optimization methods to construct low rank transition matrices $\ltp$ and $\utp$.  A low rank $\lowdim$ ensures that the lower and upper bounds to the posterior can be computed
with $O(\lowdim \statedim)$ rather than $O(\statedim^2)$ computational cost.
The transition matrices $\ltp$ and $\utp$ are constructed as low rank matrices by minimizing their {\em nuclear norms}. 
Matrices with small  nuclear norm exhibit sparseness in the set of  eigenvalues or equivalently low rank.
The  nuclear norm is the sum of the singular values of a matrix and serves as a convex surrogate of the rank of a matrix \cite{ZV09}.
 The construction of low rank transition matrices $\ltp$ and $\utp$ is formulated as a convex optimization problem on the {\em cone} of copositive matrices.\footnote{A symmetric matrix  $\mat$ is {\em  copositive} if $\belief^\p \mat \belief \geq 0$ for all positive vectors $\belief$.
(Thus the set of positive definite matrices is a subset of the set of copositive matrices. In this paper, $\belief$ are probability mass function vectors.)}
These computations are performed offline without affecting the computational cost of the real time filter.\\
3 {\em Stochastic Dominance Constrained  Monte-Carlo Importance Sampling Filter}:  
In  monitoring  systems, it is of interest to detect when the underlying Markov chain is close to a target state.
Using the reduced complexity filtering bounds outlined above, a monitoring system would want to switch to the full complexity
filter when the filtering bounds approach the target state. A natural question is:
How can the reduced complexity filtering bounds
(\ref{eq:poset}) or (\ref{eq:globalize}) be exploited to estimate the true posterior?
 Sec.\ref{sec:filters}  presents an importance sampling  Monte-Carlo method for matrix vector multiplication that is inspired by recent results in stochastic linear solvers. 
 The algorithm uses Gibbs sampling to ensure that the estimated 
posterior $\hbelief_k$ lies in the partially ordered set  $\lbelief_k \lr \hbelief_k \lr \ubelief_k$
at each time $k$.
Numerical experiments show that this stochastic dominance constrained algorithm yields estimates
with  substantially reduced mean square errors compared to the unconstrained algorithm  -- in addition, by construction
 the estimates are provably
sandwiched between $\lbelief_k$ and $\ubelief_k$.
\\
4. {\em Analytical Bounds on Variational Distance}:  Given the low complexity bounds $\lbelief_k$ and $\ubelief_k$ such that $\lbelief_k \lr \belief_k \lr \ubelief_k$, a natural question is:
How tight are the bounds? Theorem \ref{thm:copositive} 
presents explicit analytical bounds on the deviation of the true posterior  $\belief_k$ (which is expensive to compute)  from the lower and upper 
bounds $\lbelief_k$ and $\ubelief_k$ in terms of the Dobrushin coefficient of the transition matrix. It yields useful analytical bounds (that can be computed
without evaluating the posterior $\belief_k$)
for quantifying how the stochastic dominance constraints sandwich the true posterior as time evolves.

\subsection*{Related Work} The area of constructing approximate filters for estimating the state of large scale Markov chains has been well studied both in discrete and continuous time. Most works \cite{ZYM07,YZML05} assume that the Markov chain has two-time scale dynamics (e.g. the Markov chain is nearly completely decomposable). This two-time scale feature is then exploited to construct suitable filtering approximations  on the slower time scale.
In comparison, the framework in the current paper does not assume a two-time scale Markov chain. 
Indeed,
our results are finite sample results
that do not rely on asymptotics.

The main tools used in this paper are based on monotone likelihood ratio (MLR)
stochastic dominance and associated monotone structural results of the Bayesian filtering update.
Such results have been developed in the context of stochastic control and Bayesian games in \cite{Lov87,Rie91,Kri11} but have so far not been exploited to devise efficient filtering approximations.
To the best of our knowledge, constructing upper and lower sample path bounds to the optimal filter in terms of stochastic orders is new -- and the  copositivity constraints presented in this paper yield a constructive realization of these bounds. Recently,
\cite{Kri11} use similar copositive characterizations to derive structural results in stochastic control. 

Optimizing the nuclear norm as a surrogate for rank has been studied  as a convex optimization problem in several papers, see for example \cite{ZV09}.
Inspired by the seminal work of Cand{\`e}s and Tao  \cite{CT09},
there has been much recent interest in minimizing nuclear norms in the context of  sparse matrix completion problems.
  Algorithms for testing for copositive matrices and copositive programming have been studied recently in \cite{BD08,BD09}.

There has been extensive work in  signal processing on posterior Cram\'er-Rao bounds for nonlinear filtering  \cite{TMN98}; see also \cite{RAG04} for a textbook treatment.
These yield  lower bounds to the achievable variance of the conditional mean estimate of the optimal filter.
However, unlike the current paper,  such posterior Cram\'er-Rao bounds do not give  constructive algorithms for computing upper and lower bounds for the
sample path of the filtered distribution. The sample path bounds proposed in this paper have the attractive feature that 
they are guaranteed to yield lower and upper bounds to both hard and soft estimates of the optimal filter.

\section{Stochastic Dominance of Filters and Copositivity Conditions} \label{sec:copositive}
Theorem~\ref{thm:main} below is  the main result of this section -- it shows that if stochastic matrices $\ltp$ and $\utp$ are constructed such that $\ltp \lR \tp \lR \utp$ (in terms of a copositive ordering), the filtered estimates computed using $\ltp$ and $\utp$ are guaranteed to sandwich the optimal filtered estimate in terms of the monotone likelihood ratio  order.
This section sets the stage for Section \ref{sec:convex} where the construction of  low rank matrices $\ltp$ and $\utp$ is formulated as a convex optimization problem on a copositive cone; and also Section \ref{sec:filters} where algorithms that exploit this  result are presented.

\subsection{Signal Model and Optimal Filter}
Consider an $\statedim$-state discrete time Markov chain $\{\state_k\}$ on the state space $\statespace$. Suppose $\state_0 \in \statespace$ has a prior distribution  $\belief_0$. 
The $\statedim\times \statedim$-dimensional transition probability matrix $\tp$ comprises of elements $\tp_{ij} = \prob(\state_{k+1}=j | \state_k = i)$.

 The Markov process $\{\state_k\}$ is observed via a noisy process $\{\obs_k\}$ where
at each time $k$, $\obs_k \in \obspace$ or $\obs_k \in \reals^m$. As is widely assumed in optimal filtering,  we make the conditional independence assumption that  $\obs_k$ given $\state_k$ is statistically independent of $\state_{1:k-1}, \obs_{1:k-1}$.
For the case $\obs_k \in \obspace$, denote the observation likelihood probabilities as  $\oprob_{\state\obs }= \prob(\obs_k = \obs|\state_k=\state)$. 
The case  $y_k \in \reals^m$,  $\oprob_{\state\obs}$ is the conditional probability density.  (For readability to an engineering audience,
unified notation  with respect to the   Lebesgue  and counting measures is avoided.)

With  $\belief_k(i) $
denoting the posterior  defined in (\ref{eq:beliefvector}), 
the optimal filter is given by (\ref{eq:filter}).
Note that the posterior $\belief_k$ lives in an $\statedim-1$ dimensional unit simplex $\Belief$ comprising of $\statedim$-dimensional  probability vectors $\belief$. That is,
\beq  \Belief = \{\belief \in \reals^\statedim:  \one^\p \belief = 1, \quad \belief(i) \geq 0 \}. \label{eq:Belief} \eeq
Finally, since the state space  is $\statespace$, the conditional mean estimate of the state computed using the observations $\obs_{0:k}$ is (we avoid using the notation of sigma algebras)
\beq \label{eq:mean}
\mean_k \ole \E\{\state_k| \obs_{0:k};\tp\}  =  \levels^\p \state_k, \quad \text{ where } \levels = [1,2,\ldots,\statedim]^\p .\eeq
In some applications, rather than the ``soft'' state  estimate provided by the conditional mean, one is interested in the ``hard'' valued maximum aposteriori estimate defined as
\beq \label{eq:map}
  \map_k \ole \argmax_{i\in \statespace} \belief_k(i) .  \eeq

We  refer to posterior $\belief_k$ in (\ref{eq:filter}) and state estimates (\ref{eq:mean}), (\ref{eq:map}) computed using transition matrix $\tp$ as the 
``optimal filtered estimates'' to distinguish them from the lower and upper bound filters.

\subsection{Some Preliminary Definitions} \label{sec:defns}
We introduce here some key definitions that will be used in the rest of the paper.

\subsubsection{Stochastic Dominance} We start with the following standard definitions involving stochastic dominance \cite{MS02}. Recall
that $\Belief$ is the unit simplex defined in (\ref{eq:Belief}).

\begin{definition}[Monotone Likelihood Ratio (MLR) Dominance]
Let $\belief_1, \belief_2 \in \Belief$ be any two probability vectors.
Then $\belief_1$ is greater than $\belief_2$ with respect to the MLR ordering -- denoted as
$\belief_1 \gr \belief_2$ -- if
\beq \belief_1(i) \belief_2(j) \leq \belief_2(i) \belief_1(j), \quad i < j, \quad i,j\in \{1,\ldots,\statedim\}.
\label{eq:mlrorder}\eeq
\end{definition}
Similarly $\belief_1 \lr \belief_2$ if  $\leq$ in (\ref{eq:mlrorder}) is replaced
by a $\geq$. \\
The MLR stochastic order is  useful  since it is closed  under conditional expectations.
That is, $X\gr Y$  implies $\E\{X|\F\} \gr \E\{Y|\F\}$ for any two random variables $X,Y$ and sigma-algebra $\F$
\cite{Rie91,KR80,Whi82,MS02}.

\begin{definition}[First order stochastic dominance, \cite{MS02}]
 Let $\belief_1 ,\belief_2 \in \Belief$.
Then $\belief_1$ first order stochastically dominates $\belief_2$  -- denoted as
$\belief_1 \gs \belief_2$ --
 if
$\sum_{i=j}^X \belief_1(i) \geq \sum_{i=j}^X \belief_2(i)$  for $ j=1,\ldots,X$.
\end{definition}

The following result is well known \cite{MS02}. It says that MLR dominance implies first order stochastic \
dominance, and it gives a necessary and sufficient condition for stochastic dominance.
\begin{result}[\cite{MS02}] \label{res1}
 (i)  Let $\belief_1 ,\belief_2 \in \Belief$.
Then $\belief_1 \gr \belief_2$ implies $\belief_1\gs \belief_2$.\\
(ii) Let $\mathcal{V}$ denote the set of all $X$ dimensional vectors
$v$ with nondecreasing components, i.e., $v_1 \leq v_2 \leq \cdots                                                 
\leq v_X$.
Then $\belief_1 \gs \belief_2$ iff for all $v \in \mathcal{V}$,
 $v^\p \belief_1 \geq v^\p \belief_2$.
\end{result}

\begin{definition}[Total Positivity of order 2]
A transition matrix $\tp$ is totally positive of order 2 (TP2) if every second order minor of $\tp$ is non-negative. Equivalently, every row is dominated
by a subsequent row with respect to the MLR order. \label{def:tp2}
\end{definition}

\subsubsection{Copositivity} The following definitions of copositive matrices and a copositive ordering of stochastic matrices will be used extensively.
\begin{definition}[Copositivity on simplex]
An arbitrary $\statedim\times \statedim$ matrix  $\mat$ is copositive 
 if   $\belief^\p \mat \belief \geq 0$ for all $\belief \in \Belief$, or  equivalently, if  $\belief^\p \mat \belief \geq 0$ for all $\belief \in \reals^\statedim_+$.
\end{definition}
The definition says copositivity on the unit simplex and positive orthant are equivalent.
 Clearly positive semidefinite matrices and non-negative matrices are copositive.

Given two  $\statedim\times \statedim$ dimensional transition matrices $\tp$ and $Q$,  we now define a sequence of matrices
$\mat^{(m)}(Q,\tp)$, indexed by $m=1,2,\ldots,X-1$, as follows: Each  $\mat^{(m)}(Q,\tp)$ is a symmetric
 $\statedim\times \statedim$  matrix of the form:
\beq
\mat^{(m)}(Q,\tp) =  Q_m \tp_{m+1}^\p + \tp_{m+1} Q_m^\p - \tp_m Q_{m+1}^\p  - Q_{m+1} \tp_m^\p \label{eq:mat} \eeq
Here $\tp_m$ and $Q_m$, respectively, denote the $m$-th column of matrix $\tp$ and $Q$.

\begin{definition}[Copositive Ordering $\lR$ of Stochastic Matrices] \label{def:lR}
Given two  $\statedim\times \statedim$ transition matrices $\tp$ and $Q$, we say $Q \lR \tp$  (equivalently, $\tp \gR Q$) if all the matrices $\mat^{(m)}(Q,\tp)$, 
$m=1,2,\ldots,\statedim-1$,  defined in (\ref{eq:mat}), are copositive.
\end{definition}

{\em Intuition}: The  ordering of transition matrices  $Q\lR \tp$ implies that 
the optimal filtering updates satisfy $\filter(\belief,\obs;Q) \lr \filter(\belief,\obs;\tp)$ for any observation $\obs $ and posterior $\belief$.
(Recall the Bayesian update $\filter(\belief,\obs,\cdot)$ is defined in (\ref{eq:filter}) and  $\lr$ denotes the MLR order.)
In other
words the $\lR$ ordering of transition matrices preserves the MLR ordering $\lr$ of posterior distributions computed via the optimal
filter. This property will be proved in Theorem \ref{thm:main} below. This is a crucial property that will be used subsequently
in deriving lower and upper bounds to the optimal filtered posterior. It is easily verified that $\lR$ is a partial order over the set of stochastic matrices,
i.e., $\lR$ satisfies reflexivity, antisymmetry and transitivity.

\subsection{Upper and Lower Sample Path Stochastic Dominance Bounds to Posterior}

With the above definitions, we are now ready to state the main result of this section.
Recall that the original filtering problem seeks to compute $\belief_{k+1} = \filter(\belief_k,\obs_{k+1};\tp)$ using the filtering update (\ref{eq:filter}) with transition matrix $\tp$ and involves
$O(\statedim^2)$ multiplications. This can be excessive for large $\statedim$.
 Our goal is to construct low rank transition matrices $\ltp$ and $\utp$ such that the  filtering recursion using these matrices form lower and upper bounds to 
 $\belief_k$ in the MLR stochastic dominance sense. Due to the low rank of $\ltp$ and $\utp$, 
the  cost involved in computing these lower and upper bounds to $\belief_k$ at each time $k$ will be   $O(X\lowdim)$ 
 where $\lowdim \ll \statedim$ (for example, $\lowdim= O(\log \statedim)$). 

Since we plan to compute filtered estimates using $\ltp$ and $\utp$ instead of the original transition matrix $\tp$, we need  further notation to distinguish between the posteriors and estimates computed using $\tp$, $\ltp$ and $\utp$.  Let 
\begin{align*}
\underbrace{ \belief_{k+1} = \filter(\belief_k,\obs_{k+1};\tp)}_{\text{optimal}}, \quad
\underbrace{\ubelief_{k+1} = \filter(\ubelief_k,y_{k+1};\utp)}_{\text{upper}}, \\ \underbrace{\lbelief_{k+1} =  \filter(\lbelief_k,y_{k+1};\ltp)}_{\text{lower}} \end{align*}
denote the posterior  updated using optimal filter  (\ref{eq:filter}) with transition matrices $\tp$, $\utp$ and $\ltp$, respectively.
Also, similar to (\ref{eq:mean}), with $\levels = (1,2,\ldots,\statedim)^\p$, the conditional mean estimates of the underlying state computed using $\ltp$ and $\utp$, respectively, will be denoted as
\beq \label{eq:lmean}
\lmean_k \ole \E\{\state_k|\obs_{0:k};\ltp\} =  \levels^\p \lbelief_k, \quad 
\umean_k \ole \E\{\state_k|\obs_{0:k};\utp\} =  \levels^\p \ubelief_k.
\eeq
In analogy to (\ref{eq:map}), denote the ``hard" MAP state estimates computed using $\ltp$ and $\utp$  as
\beq \label{eq:lmap}
\lmap_k \ole \argmax_i \lbelief_k(i), \quad \umap_k \ole \argmax_i \ubelief_k(i).
\eeq

The following is the main result of this section. Recall the definition of copositivity ordering $\lR$, MLR dominance and TP2 in Section~\ref{sec:defns}.

\begin{framed}
\begin{theorem}[Stochastic Dominance Sample-Path Bounds] \label{thm:main}
Consider  the filtering updates $\filter(\belief,\obs;\tp)$, $\filter(\belief,y;\utp)$ and $\filter(\belief,y;\ltp)$ where $\filter(\cdot,\cdot)$ is defined in (\ref{eq:filter}) and $\tp$ denotes the transition
matrix of the original filtering problem.
\begin{compactenum}
\item For any transition matrix $\tp$, 
there exist transition  matrices 
$\ltp$ and $\utp$ such that $\ltp \lR \tp \lR \utp$ (recall $\lR$ is defined in Definition \ref{def:lR}).

\item Suppose transition matrices $\ltp$ and $\utp$ are constructed such that  $\ltp \lR \tp \lR \utp$.
%
Then  for all $\obs $ and $\belief \in \Belief$, the filtering updates satisfy the sandwich	 result
$$\filter(\belief,y;\ltp) \lr \filter(\belief,\obs;\tp) \lr \filter(\belief,y;\utp).$$
\item Suppose $\tp$  is TP2 (Definition \ref{def:tp2}).  Assume the  filters $\filter(\belief,\obs;\tp)$, $\filter(\belief,y;\utp)$ and $\filter(\belief,y;\ltp)$
 are initialized with common prior $\belief_0$ at time~$0$. Then the posteriors satisfy 
 $$ \lbelief_k \lr \belief_k \lr \ubelief_k, \quad \text{ for all time } k = 1,2,\ldots $$
 As a consequence for all time $k=1,2,\ldots$, 
 \begin{compactenum}
\item The ``soft'' conditional mean state estimates  defined in (\ref{eq:mean}), (\ref{eq:lmean}) satisfy $\lmean_k \leq \mean_k \leq \umean_k$.
\item The ``hard'' MAP state estimates defined in (\ref{eq:map}), (\ref{eq:lmap}) satisfy $\lmap_k \leq \map_k \leq \umap_k$.
\end{compactenum}
 \end{compactenum}
\end{theorem}
\end{framed}

Statement 1 says that for any transition matrix $\tp$, there always exist  transition matrices $\ltp$ and $\utp$ such that $\ltp \lR \tp \lR \utp$ (copositivity dominance).
An obvious but useless construction is $\ltp = \begin{bmatrix} e_1,\ldots, e_1 \end{bmatrix}^\p$ and
$\utp = \begin{bmatrix} e_\statedim,\ldots, e_\statedim\end{bmatrix}^\p$ where $e_i$ is the unit $\statedim$-dimensional vector with 1 in the $i$th position.
 These correspond to extreme points on the space
 of matrices with respect to copositive dominance.
Given  existence of $\ltp$ and $\utp$, the next step is to  optimize the choice of $\ltp$ and $\utp$ - that is the  subject
of Sec.\ref{sec:convex} where nuclear norm minimization is used to construct sparse eigenvalue matrices $\ltp$ and $\utp$.


Statement 2 says that  for any prior $\belief$ and observation $\obs$, the one step update of the filter lower and upper bounds
the original filtering problem.

Statement 3 globalizes Statement 2 and asserts  that with the additional assumption that the transition matrix $\tp$ of the original filtering problem is TP2, then the upper and lower bounds hold for all time.
Since MLR dominance implies first order stochastic dominance (Result \ref{res1}), the conditional mean estimates satisfy $\lmean_k \leq \mean_k \leq \umean_k$.

{\em Why MLR Dominance?}:
The proof of Theorem \ref{thm:main} in the appendix
 uses the result  that $\belief \lr \ubelief$  implies that the filtered update $\filter(\belief,\obs;\tp) \lr \filter(\ubelief,\obs;\tp)$.
Such a result does not hold
with first order stochastic dominance $\ls$ -- that is, $\belief \ls \ubelief$ does not imply that $\filter(\belief,\obs;\tp) \ls \filter(\ubelief,\obs;\tp)$.
In other words, the MLR order is closed with respect to conditional expectations. This the reason why we use the MLR order in this paper.

\subsection{Stochastic Dominance Bounds for Multivariate HMMs} \label{sec:multi}
We conclude this section
by showing how the above bounds  can be generalized to multivariate HMMs -- the main idea is that  MLR dominance is replaced
by the multivariate TP2  (totally positive of order 2) stochastic dominance \cite{MS02,Whi82,KR80}.
We consider a highly stylized example which will  serve as a reproducible way of constructing large scale HMMs in
numerical studies of Sec.\ref{sec:example}.

Consider $\nm$ independent Markov chains,
$\state_k^{(l)}$, $l=1,2\ldots, ,\nm$ with transition matrices $A^{(l)}$. 
Define the joint process
$\state_k = (\state_k^{(1)}, \ldots, ,\state_k^{(\nm)})$.
Suppose the observation process recorded at a sensor has the conditional 
probabilities $\oprob_{\state, \obs} = \prob(\obs_k = \obs| \state_k = \state)$.
Even though the individual Markov chains are independent of each other, since the observation process involves all $\nm$ Markov chains, computing the filtered estimate of $\state_k$, requires  computing and propagating
the joint posterior $P(\state_k|\obs_{1:k})$. 
 This is equivalent to HMM filtering the process $\state_k$ with transition matrix  
$\tp = A^{(1)} \otimes\cdots \otimes A^{(\nm)}$ where $\otimes$ denotes Kronecker product. For example, if each process
$\state^{(l)}$ has $S$ states, then $\tp$ is an $S^\nm \times S^\nm$ matrix and the computational cost of the HMM filter at each time is 
$O(S^{2L})$ which is excessive for large $\nm$.

A naive application of the results of the previous sections will not work, since the MLR ordering does not apply to the multivariate case. Instead, we use the totally positive  (TP2) stochastic order, which is a multivariate generalization of the MLR order.
Let $\i = (i_1,\ldots,i_\nm)$ and $\j = (j_1,\ldots,j_\nm)$ denote the indices
of two $\nm$-variate probability mass functions
Denote 
\begin{align}
\i \wedge \j &= [\min(i_1,j_1),  \ldots, \min(i_L,j_L) ]^\p, \nn \\
\i \vee \j &= [\max(i_1,j_1), \ldots, \max(i_L,j_L) ]^\p .
\end{align}

\begin{definition}[TP2 ordering and Reflexive TP2 distributions]
Let $P $ and $Q$ 
 denote any two $L$-variate probability mass functions.
Then: \\
 (i) $P \gtptwo Q$ if $P(\i) Q(\j) \leq P(\i \vee  \j) Q(\i \wedge \j)$.
 If $P$ and $Q$ are univariate, then this definition is equivalent to
 the MLR ordering  $P\gr Q$ defined above.\\
(ii)  A multivariate
distribution $P$ is said to be multivariate TP2 (MTP2) if $P \gtptwo P$  holds,
i.e.,  $P(\i) P(\j) \leq P(\i\vee  \j) P(\i \wedge \j)$.
If
 $\i,\j\in \{1,\ldots,X\}$ are scalar indices,
this is equivalent to saying that an $X\times X$ matrix $P$ is  MTP2  if all 
second order minors are non-negative.
 \label{def:tp2o}
\end{definition}
 
With suitable notational abuse,  in analogy to Definition \ref{def:lR},
given two transition matrices $\ltp$ and $\tp$ and a multivariate belief $\belief$, we say 
\beq \tp \gR \ltp ,\text { if }
\tp^\p \belief  \gtptwo \ltp^\p \belief .    \label{eq:tp2dom}\eeq
 
 The main result regarding filtering of multivariate HMMs is as follows:
 
 \begin{framed}
 \begin{theorem} \label{thm:tp2} Consider an $\nm$-variate HMM  where each transition matrix satisfies $\uA^{(l)} \lR A^{(l)}$ for $l=1,\ldots,\nm$
 (where $\lR $ is interpreted as in Definition \ref{def:lR}).
  Then\\
(i)  Then  
$\uA^{(1)} \otimes \cdots \otimes \uA^{(\nm)}  \lR  A^{(1)} \otimes \cdots \otimes A^{(\nm)} $ where $\lR$ is interpreted as (\ref{eq:tp2dom}).\\
(ii)  Theorem \ref{thm:main}  holds for the posterior and state estimates with $\gr$ replaced by $\gtptwo$.
\end{theorem}
\end{framed}
We need to qualify statement~(ii) of Theorem \ref{thm:tp2}  since 
for  multivariate HMMs,  the conditional mean $\mean_k$  and MAP estimate $\map_k$   are  $\nm$-dimensional vectors. The inequality
$\lmean_k \leq  \mean_k $ of statement (ii)  is interpreted as the component wise partial order on $\reals^\nm$, namely,  $\lmean_k(l) \leq \mean_k(l)$ for all $l=1,\ldots,\nm$. (A similar result applies for the upper bounds.)



\section{Convex Optimization to Compute Low Rank Transition Matrices $\ltp$, $\utp$} \label{sec:convex}

It only remains to 
 give algorithms for constructing  low rank transition matrices $\ltp$ and $\utp$.
 that yield the lower and upper bounds $\lbelief_k$ and $\ubelief_k$.
 These involve
  convex optimization \cite{FHB01,FHB03} for minimizing the nuclear norm.
{\em The computation of 
 $\ltp$ and $\utp$ is independent of the observation sample path and  so the associated   computational cost
 is irrelevant to the real time  filtering}. 
  Recall that the motivation is as follows:
 If $\ltp$ and $\utp$ have rank~$\lowdim$, then the computational
cost of the filtering recursion is $O(\lowdim\statedim)$ instead of $O(\statedim^2)$ at each time~$k$.

\subsection{Construction of $\ltp,\utp$ without rank constraint} \label{sec:naive}
Given a TP2 matrix $\tp$, the transition matrices $\ltp$ and $\utp$ such that
$\ltp \lR \tp \lR  \utp$ can be constructed straightforwardly via an LP solver.
With $\ltp_1,\ltp_2,\ldots,\ltp_\statedim$ denoting the rows of $\ltp$, a sufficient condition for $\ltp \lR \tp$ is that 
$\ltp_i \lr \tp_1$ for any row $i$.  
So the rows $\ltp_i$ satisfy linear constraints with respect to $\tp_1$ and can be straightforwardly constructed via an LP solver.
A similar construction holds for the upper bound $\utp$, where it is sufficient to construct $\utp_i \gr \tp_X$.

{\em Rank 1 bounds}:
If $\tp$ is TP2, 
an obvious  construction is to construct $\ltp$ and $\utp$ as follows: Choose rows $\ltp_i = \tp_1$ and $\utp_i = \tp_\statedim$ for $i=1,2,\ldots,\statedim$. These yield rank 1 matrices $\ltp$ and $\utp$.
It is clear from Theorem~\ref{thm:main} that $\ltp$ and $\utp$ constructed in this manner are the tightest rank 1 lower and upper bounds.

\subsection{Nuclear Norm Minimization Algorithms to Compute Low Rank Transition Matrices $\ltp$, $\utp$} \label{sec:alg}
 In this subsection
we construct $\ltp$ and $\utp$    as  low rank transition matrices  subject to the condition
$\ltp \lR \tp \lR \utp$. 
To save space we consider  the lower bound  transition matrix $\ltp$; construction of $\utp$ is similar.
Consider  the following   optimization problem for $\ltp$:
\begin{align}
 \text{  Minimize rank of  }  \statedim\times \statedim \text{ matrix }  \ltp   
\label{eq:obj0} 
\end{align}
subject to the constraints $\cons(\Belief,m)$ for $ m = 1,2,\ldots, \statedim-1 $, where for $\epsilon>0$,  
\begin{subequations}
\begin{empheq}[left={  \cons(\Belief,m) \equiv } \empheqlbrace]{align}
 & 
  \mat^{(m)}(\ltp,\tp)   \text{ is  copositive on $\Belief$ } \label{eq:con1} \\
&  \| \tp^\p \belief - \ltp^\p \belief \|_1 \leq \epsilon \text{ for all } \belief \in \Belief \label{eq:con2} \\
&  \ltp \geq 0 , \quad \ltp \one = \one.  \label{eq:con3} \end{empheq} \\
\end{subequations}
Recall  $\mat$ is defined in (\ref{eq:mat}).
The constraints $\cons(\Belief,m)$ are convex in matrix $\ltp$, since  (\ref{eq:con1}) is linear in the elements of $\ltp$ and  (\ref{eq:con2}) is convex because
 norms are convex.
The constraints (\ref{eq:con1}), (\ref{eq:con3}) are exactly the conditions of Theorem \ref{thm:main}.
Recall that (\ref{eq:con1}) is equivalent to $ \ltp \lR \tp$.
The  convex constraint (\ref{eq:con2}) is equivalent to
$\|\ltp-\tp\|_1 \leq \epsilon$, where $\|\cdot\|_1$ denotes the induced 1-norm for matrices.\footnote{
The three  statements  $\| \tp^\p \belief - \ltp^\p \belief \|_1 \leq \epsilon$,  $\|\ltp-\tp\|_1 \leq \epsilon$ and $\sum_{i=1}^\statedim \| (\tp^\p - \ltp^\p)_{:,i} \|_1 \belief(i)  \leq \epsilon$
are all equivalent since $\|\belief\|_1 = 1$.}

%
%


To solve the above problem, we proceed in two steps:
\begin{compactenum}
\item The objective (\ref{eq:obj0}) is replaced with the reweighted nuclear norm (Sec.\ref{sec:nuclear} below).
\item
Optimization over the copositive cone (\ref{eq:con1}) is achieved via a sequence of simplicial decompositions (Sec. \ref{sec:simplex} below).
\end{compactenum}

\subsubsection{Reweighted Nuclear Norm} \label{sec:nuclear}
Since 
  the	 rank	   is a non-convex function of a matrix,  direct minimization of the rank (\ref{eq:obj0}) is  computationally intractable.
Instead,  we follow the approach developed by Boyd and coworkers \cite{FHB01,FHB03} to minimize the  iteratively reweighted  nuclear norm.
As mentioned earlier, inspired by  Cand{\`e}s and Tao \cite{CT09}, there has been much recent interest in minimizing nuclear norms for constructing matrices with sparse eigenvalue sets
or equivalently low rank. Here we compute $\ltp, \utp$ by
minimizing their nuclear norms subject to copositivity conditions that ensure $\ltp \lR \tp \lR \utp$. 

The re-weighted nuclear norm minimization
 proceeds as a {\em sequence} of convex optimization problems indexed by $n=0,1,\ldots$.
Initialize $\ltp^{(0)} = I$. For $n=0,1,\ldots$, compute $\statedim \times \statedim$ matrix
\begin{align}
\ltp^{(n+1)} &= \argmin_{\ltp} \| \underline{W}_1^{(n)} \ltp \; \underline{W}_2^{(n)} \|_*   \label{eq:obj} \\
\text{ subject to: } & \text{ constraints  $\cons(\Belief,m)$, $m=1,\ldots,\statedim-1$ }  \nn \\ & \text{ namely,  (\ref{eq:con1}), (\ref{eq:con2}), (\ref{eq:con3}). } \nonumber
\end{align}
Here $\| \cdot \|_*$ denotes the nuclear norm, which corresponds to the sum of the singular values of a matrix, and the weighting matrices $\underline{W}_1^{(n)}$, $\underline{W}_2^{(n)}$ are evaluated iteratively as
\begin{align}
\underline{W}_1^{(n + 1)} &= ([\underline{W}_1^{(n)}]^{-1} U \Sigma U^T [\underline{W}_1^{(n)}]^{-1} + \delta I)^{-1/2}, \nn
\\
\underline{W}_2^{(n + 1)} &= ([\underline{W}_2^{(n)}]^{-1} V \Sigma V^T [\underline{W}_2^{(n)}]^{-1} + \delta I)^{-1/2}. \label{eq:w1w2}\end{align}
Here $\underline{W}_1^{(n)} \ltp^{(n)} \underline{W}_2^{(n)} = U \Sigma V^T$ is a reduced singular value decomposition, starting with $\underline{W}_1^{(0)} = \underline{W}_2^{(0)} = I$ and $\ltp^{0} = \tp$. Also $\delta$ is a small positive constant  in the regularization term $\delta I$.  
In numerical examples of Sec.\ref{sec:example}, we used {\tt YALMIP} with {\tt MOSEK} and {\tt CVX}  to solve the above convex optimization problem.

Let us explain the above sequence of convex optimization problems.
Notice that at iteration $n+1$, the previous estimate, $\ltp^{(n)}$ appears in  the cost function of \eqref{eq:obj} in terms of
weighing matrices
 $\underline{W}_1^{(n)}$, $\underline{W}_2^{(n)}$. The intuition behind the reweighing iterations is that as the estimates $\ltp^{(n)}$ converge to the limit $\ltp^{(\infty)}$, the cost function becomes approximately equal to the rank of $\ltp^{(\infty)}$.

\subsubsection{Simplicial Decomposition for copositive programming} \label{sec:simplex}

Problem \eqref{eq:obj} is  a convex optimization problem in $\ltp$.
 However, one additional issue needs to be resolved: the constraints \eqref{eq:con1} involve a copositive cone and cannot be solved directly by standard interior point methods.  
To deal with the copositive constraints  \eqref{eq:con1}, we use the  state-of-the-art simplicial decomposition
method proposed in~\cite{BD09}.  The nice key idea used in \cite{BD09} is summarized in the following proposition.
\begin{proposition}[\cite{BD09}] Let $\Lambda $  denote any sub-simplex of the belief space $\Belief$. Then a sufficient condition for copositive condition (\ref{eq:con1})  to hold on $\Lambda$ is that  it
 holds on the  vertices of $\Lambda$.  \label{prop:band}
\end{proposition}

Let $\Lambda_j \ole \{\Lambda^{J}_1,\ldots,\Lambda^J_{L}\}$ denote the set of subsimplices at iteration $J$ that constitute a partition of $\Belief$.
Proposition \ref{prop:band} along with the nuclear norm minimization leads to a finite dimensional convex optimization problem that can be solved via the following  2 step  algorithm:  

\noindent  {\bf for}  iterations $J=1,2\ldots$,
\begin{compactenum}
\item  Solve the sequence of convex optimization problems (\ref{eq:obj}), $n=1,2,\ldots$ 
with constraints $\cons(\Lambda^J_1, m), \cons(\Lambda^J_2, m), \ldots, \cons(\Lambda^J_L, m)$,  $m =1 ,\ldots,X-1$.
\item {\bf if}  nuclear norm $\|  \underline{W}_1^{(n)} \ltp \; \underline{W}_2^{(n)} \|_*$ decreases compared to that in iteration $J-1$ by more than a pre-defined  tolerance, systematically partition $\Lambda_J$ as described in \cite{BD09}
into $\Lambda_{J+1}$. \\ Set $J = J+1$ and go to Step 1.\\
{\bf else} Stop.
\end{compactenum}
The iterations of the above simplicial algorithm lead to a sequence of decreasing costs $\|  \underline{W}_1^{(n)} \ltp \; \underline{W}_2^{(n)} \|_*$, hence the algorithm can be terminated as soon as the decrease in the cost becomes smaller than a pre-defined value  (set by the user); please see \cite{BD09} for details on simplex partitioning.
We emphasize again that the algorithms in this section for computing $\ltp$ and $\utp$ are off-line and do not affect the real time filtering computations.

\section{Stochastic Dominance Constrained Importance Sampling Filter} \label{sec:filters}
So far we have 
 constructed reduced complexity
 lower and upper  stochastic dominance bounds that confine  the posterior  sample path
of the optimal filter to the partially ordered set
$\lbelief_k \lr \belief_k \lr \ubelief_k$ at each  time $k$. 
The next question is: Given the  estimates $\lbelief_k$ and $\ubelief_k$, how to construct an algorithm to estimate $\belief_k$?
That is, how can the bounds $\lbelief_k$ and $\ubelief_k$ be exploited to estimate the posterior $\belief_k$?
We present  a filtering algorithm that is inspired by recent  results
in stochastic linear solvers \cite{ESS11,DKM06}. The algorithm
uses  importance sampling for matrix-vector multiplication together with Gibbs sampling to ensure that   the estimated posterior $\hbelief_k$ 
lies in the partially ordered set $\lbelief_k \lr \hbelief_k \lr \ubelief_k$.  

{\em Why?} Running a reduced complexity estimator and then switching to a high resolution estimator when an event of  interest occurs,
arises in 
monitoring systems, cued sensing in adaptive target tracking  systems \cite{BP99} and  body area networks \cite{BPH05,PBC03}.
In these examples,
 it is of interest to detect when the underlying Markov chain is close to a target state.
A sensor monitoring the state of a noisy Markov chain can compute the reduced 
complexity filtering
bounds cheaply. Since the reduced complexity bounds provably sandwich the true posterior, as soon as these bounds get close to a target state, the sensor switches to a higher resolution (complexity) estimator.
For example, in cued target tracking, when a target's state approaches a high threat level, the reduced complexity tracker
can cue  (deploy) a higher resolution (complexity) tracker.

{\em Remark}: We emphasize at the outset that  obviously, $P(\state_k | \obs_{1:k}) = P(\state_k | \obs_{1:k}, \lbelief_k, \ubelief_k)$
since $\lbelief_k$ and $\ubelief_k$ are $\obs_{1:k}$ measurable.
 That is, the posterior (and therefore, the conditional mean
estimate)
is exactly the same
whether or not the upper and lower bounds are used. (In other words, since the upper and lower bounds were computed using the same observations as the conditional mean estimate, they cannot be used to obtain a better
conditional mean estimate.)  This section deals with {\em estimating} the posterior - the posterior estimate conditioned on the upper and lower bounds has a lower variance
than the unconditional estimator.

\subsection{Stochastic Dominance Constrained Importance Sampling Filtering Algorithm} \label{sec:importance}



  Suppose we have an estimate $\hbelief_{k-1}$ of the posterior  such that $\lbelief_{k-1} \lr \hbelief_{k-1} \lr \ubelief_{k-1}$.
Algorithm~\ref{alg:pf} constructs an estimate $\hbelief_{k|k-1}$  using Monte-Carlo important sampling methods for matrix-vector multiplication so that 
 the predicted distributions satisfy $ \lbelief_{k| k-1} \lr  \hbelief_{k|k-1} \lr  \belief_{k|k-1}$.
Once $ \lbelief_{k| k-1} $ is constructed, the filtered posterior at time $k$ is straightforwardly computed  with  $O(\statedim)$ computations as
$ \hbelief_{k}  \propto  \oprob_{\obs_{k}}\, \hbelief_{k| k-1} $ (Bayes rule). Moreover, by Theorem  \ref{thm:main}, this updated posterior is
guaranteed to satisfy
$\lbelief_{k} \lr \hbelief_{k} \lr \ubelief_{k}$.


\begin{algorithm}\caption{Stochastic Dominance Constrained Importance Sampling Filter at time $k$} \label{alg:pf}
\begin{algorithmic}
\STATE {\bf Aim:} Given  posterior estimate $\hbelief_{k-1}$, lower bound    $\lbelief_{k-1}$ and upper bound $\ubelief_{k-1}$, evaluate
 $\hbelief_k$.
 \STATE {\bf Step 0 (offline)}: Given TP2 transition matrix $\tp$, compute low rank $\ltp$ and $\utp$ with $\ltp \lR \tp \lR \utp$
 by minimizing nuclear norm (Sec.\ref{sec:alg}). 
 \STATE  {\bf Step 1}:  Evaluate   predicted \& filtered upper/lower bounds 
\begin{align}  \lbelief_{k|k-1} = \ltp^\p \lbelief_{k-1}, \quad \lbelief_{k} = \frac{\oprob_{\obs_{k}} \lbelief_{k|k-1}}{\one^\p\oprob_{\obs_{k}} \lbelief_{k|k-1}} \nn \\
\ubelief_{k|k-1}= \utp^\p \ubelief_{k-1}, 
\quad
 \ubelief_{k} = \frac{\oprob_{\obs_{k}} \ubelief_{k|k-1}}{\one^\p\oprob_{\obs_{k}} \ubelief_{k|k-1}}
 \label{eq:rcfb}
\end{align}
\STATE {\bf Step 2}: Compute estimate $\hbelief_k$ using $\lbelief_k$ and $\ubelief_k$.
\FOR { $j=1$ \TO $\statedim$ }
   \STATE Evaluate  stochastic dominance  path bounds $\lb_j$ and $\ub_j$:  
    $\lb_1 = 0$, $\ub_1= 1$ and for $j>1$, 
   \begin{align}  \lb_j &=\frac {  \lbelief_{k|k-1}(j)\,   \hbelief_{k|k-1}(j-1)}{\lbelief_{k|k-1}(j-1)} ,  \label{eq:lbub} \\  \ub_j &= \min\left\{\frac {  \ubelief_{k|k-1}(j)\,  \hbelief_{k|k-1}(j-1)}{\ubelief_{k|k-1}(j-1)}    ,
1 - \sum_{l=1}^{j-1}  \ubelief_{k|k-1}(l)  \right\}. \nn
\end{align}
\FOR { iterations $\iter = 1$ \TO $\Iter$ } \STATE Sample $i_\iter \in \{1,\ldots,\statedim\}$ from importance probability mass function $\impdensity_j$. \IF{ $\frac{\tp_{i_\iter,j}\, \belief_{k-1}(i_\iter)}{\impdensity(i_\iter)} \in [\lb_j,\ub_j]$}
\STATE 
\beq \text{ Set $F_j = F_j \cup \{l\}$ and }
 \hbelief^{(l)}_{k|k-1}(j) =  \frac{\tp_{i_{\iter},j}\, \hbelief_{k-1}(i_{\iter})  }{\impdensity_j(i_{\iter}) }  \label{eq:pfilter} \eeq
%
\ENDIF
\ENDFOR
\STATE $ \text{ Set }  \hbelief_{k|k-1}(j) = \frac{1}{|F_j|} \sum_{l \in F_j}  \hbelief^{(l)}_{k|k-1}(j) $.  \\ (If $F_j$ is empty, set $ \hbelief_{k|k-1}(j)  =
\lbelief_{k|k-1}(j)$)
\ENDFOR
\STATE Compute filtered posterior estimate $\hbelief_{k} =  \frac{\oprob_{\obs_{k}} \hbelief_{k|k-1} }{ \one^\p  \oprob_{\obs_{k}} \hbelief_{k|k-1}} $
\end{algorithmic}
\end{algorithm}


Eq.(\ref{eq:lbub}) in Algorithm  \ref{alg:pf} is equivalent to
$\lb_j \leq \hbelief_{k|k-1}(j) \leq \frac {  \ubelief_{k|k-1}(j)\,  \hbelief_{k|k-1}(j-1)}{\ubelief_{k|k-1}(j-1)}  $. This  in turn is equivalent to the sample path bound $\lbelief_{k|k-1} \lr \hbelief_{k|k-1} \lr \ubelief_{k|k-1}$.
 The key point in Algorithm \ref{alg:pf} is the reduced variance compared to the 
 un-constrained estimator
 since  $\text{var}( \hbelief | \lbelief \lr \hbelief \lr \ubelief )  \leq \text{var}(\hbelief)$.
If the  stochastic dominance constraints are not exploited, then
$\lb_j = 0$ and $\ub_j =  1 - \sum_{l=1}^{j-1}  \ubelief_{k+1|k}(l)$ in (\ref{eq:lbub}).
The condition (\ref{eq:pfilter}) uses Gibbs sampling  to ensure that the constraints hold - this is simply a special case of adaptive importance sampling.\footnote{We thank Eric Moulines of ENST for mentioning this.}

 {\em Choice of Importance Distribution}:
In Algorithm \ref{alg:pf}, the importance distribution $\impdensity_j$ is an $\statedim$-dimension probability vector.
There are several choices for the importance distribution $\impdensity_j$.
\begin{compactenum}
\item 
An obvious  choice is $\impdensity_j(i) = \hbelief_k(i)$, in  which case  (\ref{eq:pfilter}) becomes:  If  $\tp_{i_\iter,j}   \in [\lb_j,\ub_j]$ then 
$ \hbelief_{k|k-1}(j) \gets   \tp_{i_{\iter+1},j} $
\item
The optimal importance function, which minimizes the variance of $\hbelief_{k|k-1}$, is $\impdensity_j(i) \propto  \tp_{ij}\, \hbelief_{k-1}(i)$. 
This is not useful since evaluating it requires $O(\statedim)$ multiplications for each $j$ and therefore $O(\statedim^2)$ multiplications in total.
\item A near optimal choice is to choose  $\impdensity_j(i) \propto  \ltp_{ij} \hbelief_{k-1}(i)$ or  $\impdensity_j(i) \propto  \utp_{ij} \hbelief_{k-1}(i)$.
These have already been computed and therefore no extra computations are required.
 These are particularly useful
when $\ltp$ and $\utp$ are constructed to minimize the distance between the bounds and the actual posterior (as discussed in Sec.\ref{sec:alg} below).
\end{compactenum}
One can add an 
optional step below (\ref{eq:pfilter}) to increase the sampling efficiency - if a particular index $i_\iter$ does not satisfy the constraint, then there is no need to simulate it again;
simulation of this index it can be eliminated by setting the corresponding probability $\impdensity_j(i_\iter) = 0$.

(iii) {\em Algorithm \ref{alg:pf} is not a particle filter}.
Algorithm  \ref{alg:pf}, in particular, (\ref{eq:pfilter}), is simply a Monte-Carlo evaluation of the matrix multiplication $\tp^\p \hbelief_{k-1}$ and is motivated by techniques in \cite{DKM06,ESS11}. In particular,
(\ref{eq:pfilter}) without the constraints, is simply Algorithm 1 of \cite{ESS11}. 
Degeneracy issues that plague particle filtering do not arise.   For $L$ iterations at each time instant $k$, Algorithm   \ref{alg:pf} has $O(\statedim ( \Iter+ \lowdim))$ computational cost where $\lowdim$ is the rank
of $\ltp$.  In comparison a particle filter with $L$ particles involves $O(L)$ computational
cost.

\subsection{Importance Sampling Filter for Computing Lower Bound}

Given the lower bound matrix $\ltp$ of rank $\lowdim$,
 $\lbelief_k$ can be computed exactly using (\ref{eq:rcfb}) with $O(\lowdim \statedim)$ computations.
An alternative method is to exploit the rank $\lowdim$ and estimate $\lbelief_k$ by using
Monte-Carlo importance sampling methods similar to Algorithm \ref{alg:pf}.  Consider the singular value decomposition of $\ltp$:
\beq
\ltp^\p \belief =  \sum_{r=1}^\lowdim  \sv_r  v_r  u_r^\p \belief 
\eeq
where we have minimized rank $\lowdim$ via the nuclear norm minimization algorithm of Sec.\ref{sec:alg}.
 Algorithm \ref{alg:upf} presents the importance sampling filter for $\lbelief$ (the upper bound is similar).

\begin{algorithm}\caption{Importance Sampling Filter for estimating lower bound $\lbelief_k$ at time $k$} \label{alg:upf}
\begin{algorithmic}
\STATE {\bf Aim:} Given  lower bound  estimate $\hlbelief_{k-1}$,  evaluate lower bound
 $\hlbelief_k$.
%
\FOR{ $r = 1$ to $\lowdim$}
\FOR { iterations $\iter = 1$ \TO $\Iter$ } \STATE Sample $i_\iter \in \{1,\ldots,\statedim\}$ from importance probability mass function $\impdensity_r$. 
\STATE Set $ \hat{u}_r(l)  = \frac{u_r(\iter)\, \hlbelief_{k-1}(i_\iter)}{\impdensity_r(i_\iter)} $
%
\ENDFOR
\STATE Set $\hat{u}_r = \frac{1}{\Iter}\sum_{\iter=1} ^\Iter  \hat{u}_r(l) $
\ENDFOR
\STATE  Set   $\hlbelief_{k|k-1} =  \sum_{r=1}^\lowdim  \sv_r v_r  \hat{u_r}   $  (where the  vectors
$\sigma_r v_r$, $r=1,\ldots,\lowdim$ are precomputed).
\STATE Compute filtered posterior $\hbelief_{k} =  \frac{\oprob_{\obs_{k}} \hlbelief_{k|k-1} }{ \one^\p  \oprob_{\obs_{k}} \hlbelief_{k|k-1}} $
\end{algorithmic}
\end{algorithm}
The choice of importance sampling distributions $\impdensity$ is similar to that for Algorithm \ref{alg:pf}.

\subsection{Stochastic Dominance Constrained Particle Filter -- A Non-result} \label{sec:particle}
Given the abundance of publications in particle filtering, it is of interest to obtain a particle filtering algorithm that exploits the upper and lower bound
constraints to estimate the posterior. Unfortunately, since particle filters propagate trajectories and not marginals,
we were unable to find a computationally efficient way of enforcing the MLR constraints
$\lbelief_k \lr \hbelief_k \lr \ubelief_k$ in the computation of $\hbelief_k$.  (If we propagated the marginals, then the algorithm becomes identical to
Algorithm \ref{alg:pf}.) Also, since MLR comparison of two $X$-dimensional posteriors involves
$O(X)$ multiplications, projecting  $L$ particles to the polytope $\lbelief_k \lr \hbelief_k\lr  \ubelief_k$ involves $O(L\statedim)$ computations.
Finally, in the particle filtering folklore, the so called `optimal' choice for the  importance density  is $\impdensity(\state_k|\state_{0:k-1},\obs_{1:k}) = \prob( \state_k | \state_{k-1},\obs_k) $ with particle weight update
$w_k^{(l)} = w_{k-1}^{(l)}  \sum_{j=1}^\statedim \tp_{\state_{k-1}^{(l)} j} \oprob_{j\obs_k}$. For each particle, this requires $O(\statedim)$ computations and hence $O(LX)$ for $L$ particles.

\section{Analysis of Bounds}

Our main result, namely, 
Theorem \ref{thm:main}  above,  is an {\em ordinal} bound: It  said that we can compute reduced complexity filters $\lbelief_k$ and $\ubelief_k$ such
that  the posterior $\belief_k$ of the original filtering problem is lower and upper  bounded on the partially ordered set:
$\lbelief_k \lr \belief_k \lr \ubelief_k$ for all time $k$. 
Moreover, by minimizing  (\ref{eq:obj}), we computed transition matrices $\ltp$ and $\utp$ so
that  $ \| \tp^\p \belief - \ltp^\p \belief \|_1 \leq \epsilon $  and 
 $ \| \tp^\p \belief - \utp^\p \belief \|_1 \leq \epsilon $.


In this section we construct {\em cardinal} bounds -- that is,
 an explicit analytical bound is developed for 
$\|\lbelief_k - \belief_k\|$ and therefore $|\lmean_k - \mean_k|$ in terms of $\epsilon$.
These bounds together with Theorem \ref{thm:main} give a complete characterization of the reduced
complexity filters.

In order to present the main result, we first define the Dobrushin coefficient:

\begin{definition}[Dobrushin Coefficient]  
For a transition matrix $\tp$,   the Dobrushin coefficient of ergodicity is
\beq \label{eq:dob}
\dob(\tp) =  \frac{1}{2}  \max_{i,j}  \sum_{l\in \statespace} | \tp_{il}- \tp_{jl} | .
\eeq
\end{definition}
Note that $\dob(\tp)$ lies in the interval $ [0,1]$. Also $\dob(\tp) = 0$ implies that the process $\{\state_k\}$ is independent and identically distributed (iid).
In words: the Dobrushin coefficient of ergodicity $\dob(\tp)$ is the maximum variational norm\footnote{It is conventional
to use the variation norm  to measure the distance between two probability distributions.
Recall that given  probability mass functions $\alpha$ and $\beta$ on $\statespace$,  the variational norm is 
$\dvar{\alpha}{\beta}=  \frac{1}{2} \|\alpha- \beta\|_1 = \frac{1}{2} \sum_{i \in \statespace}| \alpha(i) -
\beta(i)|$.
So the variational norm is just half the $l_1$ norm  between two probability mass functions.} between
two rows of the transition matrix $\tp$.


The following is the main result of this section:

\begin{framed}
\begin{theorem} \label{thm:copositive} Consider a HMM with transition matrix $\tp$ and state levels $\levels$.
Let $\epsilon > 0$ denote the  user defined parameter  in constraint (\ref{eq:con2}) of convex optimization
problem (\ref{eq:obj}) and let $\ltp$ denote the solution.  
Then
\begin{compactenum}
\item The expected absolute deviation between one step of filtering using $\tp$ versus $\ltp$ is upper bounded as:
\begin{multline} \E_{\obs}  \left| \levels^\p \left( \filter(\belief,\obs;\tp) - \filter(\belief,\obs;\ltp) \right) \right| \leq \\
\epsilon \sum_\obs \max_{i,j}  \levels^\p (I - \filter(\belief,\obs;\ltp) \one^\p) \oprob_\obs (e_i - e_j)
\label{eq:ebound}
\end{multline}

\item The sample paths of the filtered posteriors and conditional means have the  following explicit bounds at each time $k$:
\begin{multline} \label{eq:samplepath1}
\| \belief_k - \lbelief_k\|_1 \leq \frac{\epsilon}{ \max\{ \statem(\lbelief_{k-1},\obs_k) - \epsilon, \, \mu(\obs_k) \}}  \\  + \frac{ \dob(\ltp) \, \|\belief_{k-1} - \lbelief_{k-1} \|_1}{\statem(\lbelief_{k-1},\obs_k)}
%
\end{multline}
Here $\dob(\ltp)$ denotes the Dobrushin coefficient of the transition matrix $\ltp$ and   $\lbelief_k$ is  the posterior
computed using the HMM filter with  $\ltp$, and
\beq \label{eq:mudef}
\statem(\lbelief,\obs) = \frac{\one^\p \oprob_{\obs} \ltp^\p \lbelief}{\max_i \oprob_{i,\obs}}, \quad
\mu(\obs) = \frac{ \min_i \oprob_{iy} } { \max_i \oprob_{iy} } .
\eeq
\end{compactenum}
\end{theorem} \end{framed}

 Theorem \ref{thm:copositive} gives explicit upper bounds between the filtered distributions using 
 transition matrices $\ltp$ and $\utp$. The $\E_\obs$ in (\ref{eq:ebound}) is with respect to the 
 measure $\filterd(\belief,\obs;\tp) = \one^\p \oprob_y \tp \belief$  which corresponds to $\prob(\obs_k=\obs| \belief_{k-1} = \belief)$. Similar bounds hold for $\utp$ and are omitted.
 
The bounds are  useful since  their
computation  involves the reduced complexity filter with transition matrices $\ltp$  -- the original transition matrix $\tp$ is not used.
 In numerical examples below, we illustrate (\ref{eq:ebound}).

\section{Numerical examples} \label{sec:example}
In this section we present  numerical examples to illustrate the behavior of the reduced complexity filtering algorithms proposed in this paper. To give the reader an easily reproducible numerical example of large dimension,  we construct a 3125 state Markov chain according to the multivariate HMM
construction detailed in Sec.\ref{sec:multi}.
Consider  $\nm= 5$ independent Markov chains $x_k^{(l)}$, $l=1,\ldots,5$,  each with  5 states.
The observation process is 
$$ \obs_ k = \sum_{l=1}^5 x_k^{(l)} + \onoise_k $$
where the observation noise  $\onoise_k$ is zero mean iid Gaussian with variance $\onoisevar$.
Since the observation process involves all 5 Markov chains,
computing the filtered estimate requires propagating the joint posterior. 
This is equivalent to defining a $5^5=3125$ state Markov chain with transition matrix 
$  \tp = A \underbrace{\otimes\cdots \otimes}_{5 \text{ times}} A$ where $\otimes$ denotes Kronecker product.
The optimal HMM  filter incurs  $5^{10} \approx 10$ million  computations at each time step $k$.

\subsubsection{Generating TP2 Transition Matrix}
To illustrate the reduced complexity global sample path bounds developed in Theorem  \ref{thm:main},  we consider the case where $\tp$ is TP2. 
We used the following  approach to generate $\tp$: First construct   $A = \exp(Q t)$, where $Q$ is a tridiagonal generator 
matrix (nonnegative off-diagonal entries and each  row adds to $0$) and $t>0$.
 Karlin's classic book  \cite[pp.154]{KT81} shows that $A$ is then TP2.
Second, as shown in  \cite{KR80},  the Kronecker products of $A$ preserve the TP2 property 
implying that $\tp$ is TP2.

Using the above procedure, we constructed a $3125\times 3125$ TP2 transition matrix $\tp$ as follows:
\begin{align}  Q &= \begin{bmatrix}  -0.8147   &   0.8147  &        0 &        0 &        0 \\
    0.4529   &  -0.5164  &     0.0635       0   &      0 \\
         0 &    0.4567  &  -0.7729 &     0.3162 &      0 \\  
         0   &       0    &   0.0488  &  -0.1880  & 0.1392   \\
         0   &       0   &       0   &    0.5469 &  -0.5469  \end{bmatrix},\nn  \\ A &= \exp(2 Q), \quad   \tp = A \underbrace{\otimes\cdots \otimes}_{5 \text{ times}} A.
         \label{eq:tp5}
\end{align}

\subsubsection{Off-line Optimization of  Lower Bound via Convex Optimization} 
We used the semidefinite optimization {\tt solvesdp} solver from {\tt MOSEK} with  {\tt YALMIP} and {\tt CVX} to solve the convex optimization problem  (\ref{eq:obj})  for computing the upper and lower
bound transition matrices $\ltp$ and $\utp$. To estimate the rank of the resulting transition matrices, we consider the costs \eqref{eq:obj}, which correspond approximately to the number of singular values larger than $\delta$ (defined in (\ref{eq:w1w2})). The reweighed nuclear norm algorithm is run for $5$ iterations, and the simplicial algorithm is stopped as soon as the cost decreased by less than $0.01$.

To save space we  present results only for the lower bounds.
We computed\footnote{In each case, after computing the low rank matrix $\ltp$, small   singular
values of $\ltp$ were truncated to zero. The resulting matrix was then made stochastic by subtracting the
minimum element of the matrix (thereby every element is non-negative) and then normalizing the rows.
Both transformations do not affect the rank of the matrix. It was ensured that the resulting matrix $\hat{\ltp}$ satisfies the normalized error bound
$\frac{  \| \ltp^\p \belief -  \hat{\ltp} \belief \|_2  } {\| \ltp\|_2  \|\belief\|_2 } \leq  
\frac{ \| \hat{\ltp} - \ltp\|_2 }{ \| \ltp \|_2 } \leq  0.01$, thereby implying that approximating $\ltp$ by $\hat{\ltp}$ results in negligible error. For notational convenience, we continue
to use $\ltp$ instead of $\hat{\ltp}$. } 
5 different lower bound transition matrices $\ltp$ by solving the nuclear norm minimization problem (\ref{eq:obj}) for 5 different choices of $\epsilon\in \{0.4,0.8,1.2,1.6,2\}$ defined
in constraint (\ref{eq:con2}).

\begin{table} \centering
\begin{tabular}{|c|c|c|c|c|c|c|}
\hline
$\epsilon $ &  0 &   0.4 & 0.8 & 1.2 & 1.6 & 2 \\
\hline
$\lowdim $ (rank of $\ltp$) & 3125 ($\ltp = \tp$)&  800 &  232 & 165  & 40 &  1 (iid) \\ \hline
\end{tabular}
\caption{Ranks of lower bound transition  matrices $\ltp$ each of dimension $3125 \times 3125$ obtained as solutions of the nuclear norm minimization problem (\ref{eq:obj}) for six
  different choices of $\epsilon$ appearing in constraint (\ref{eq:con2}). Note $\epsilon = 0$ corresponds to $\ltp = \tp$ and $\epsilon = 2$ corresponds to the iid case.}
\label{tab:ltp}
\end{table}

\begin{figure}
\centering
\includegraphics[width=0.7\columnwidth]{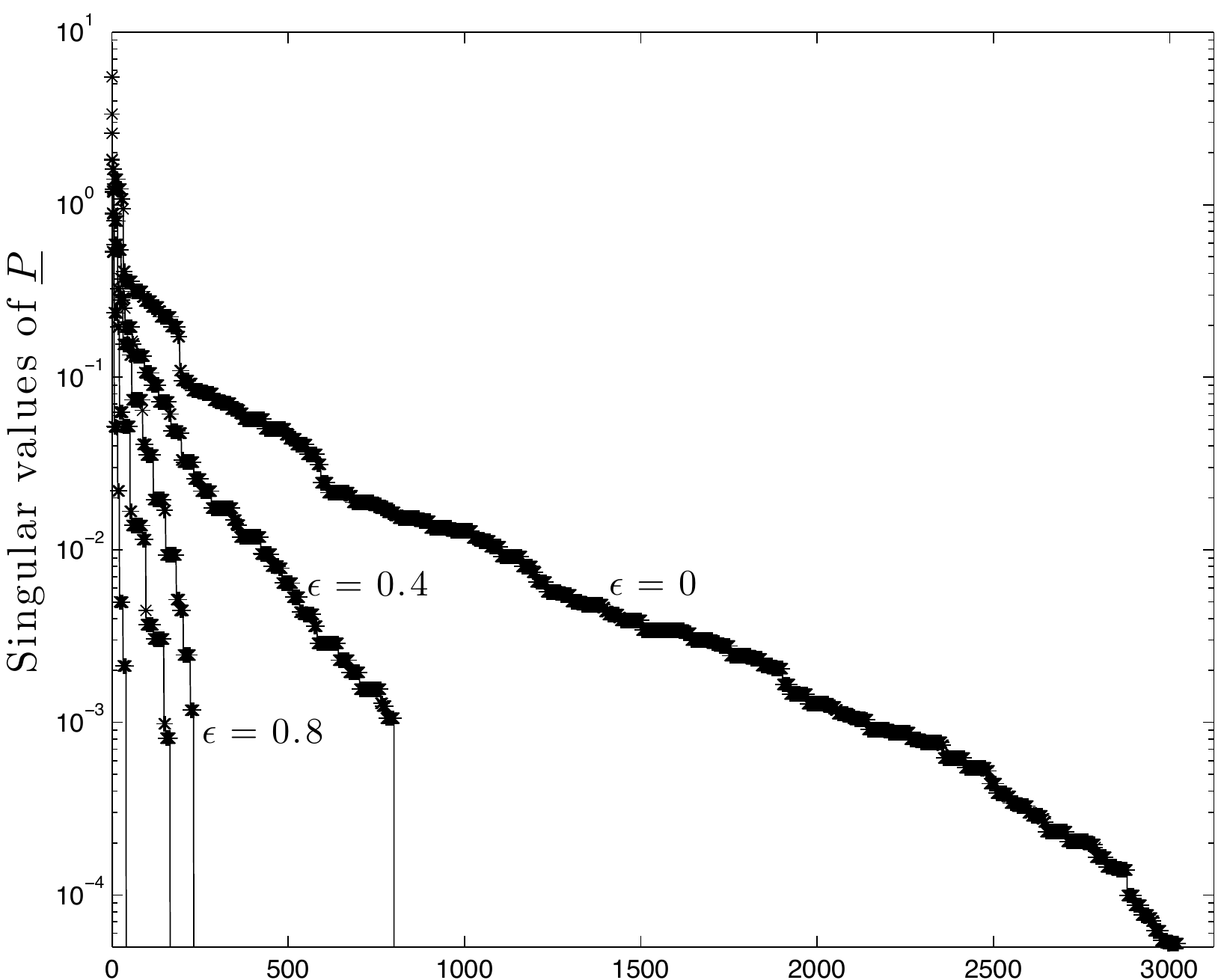}
\caption{Plot of 3125 singular values of $\tp$ and singular values of five different transition  matrices $\ltp$ parametrized by $\epsilon$ in 
Table \ref{tab:ltp}. The transition matrix $\tp$ (corresponding to $\epsilon =0$) of dimension $3125 \times 3125 $ is specified in (\ref{eq:tp5}). }
\label{fig:svd}
\end{figure}

Table \ref{tab:ltp} displays the ranks of these 5 transition matrices $\ltp$, and also the rank of $\tp$ which corresponds to the case $\epsilon = 0$.
The low rank property of $\ltp$ can be visualized by displaying the singular values.
Fig.\ref{fig:svd} displays the singular values of $\ltp$ and  $\tp$.   When $\epsilon = 2$,  the rank of $\ltp$ is  1 and models an iid chain; $\ltp$ then simply comprises of repetitions of the first row 
of $\tp$.   As $\epsilon$ is made smaller the number of singular values increases. 
For $\epsilon = 0$, $\ltp$ coincides with $\tp$.

\subsubsection{Performance of Lower Complexity Filters} \label{sec:lcf}
\begin{figure}
\centering
\includegraphics[width=0.7\columnwidth]{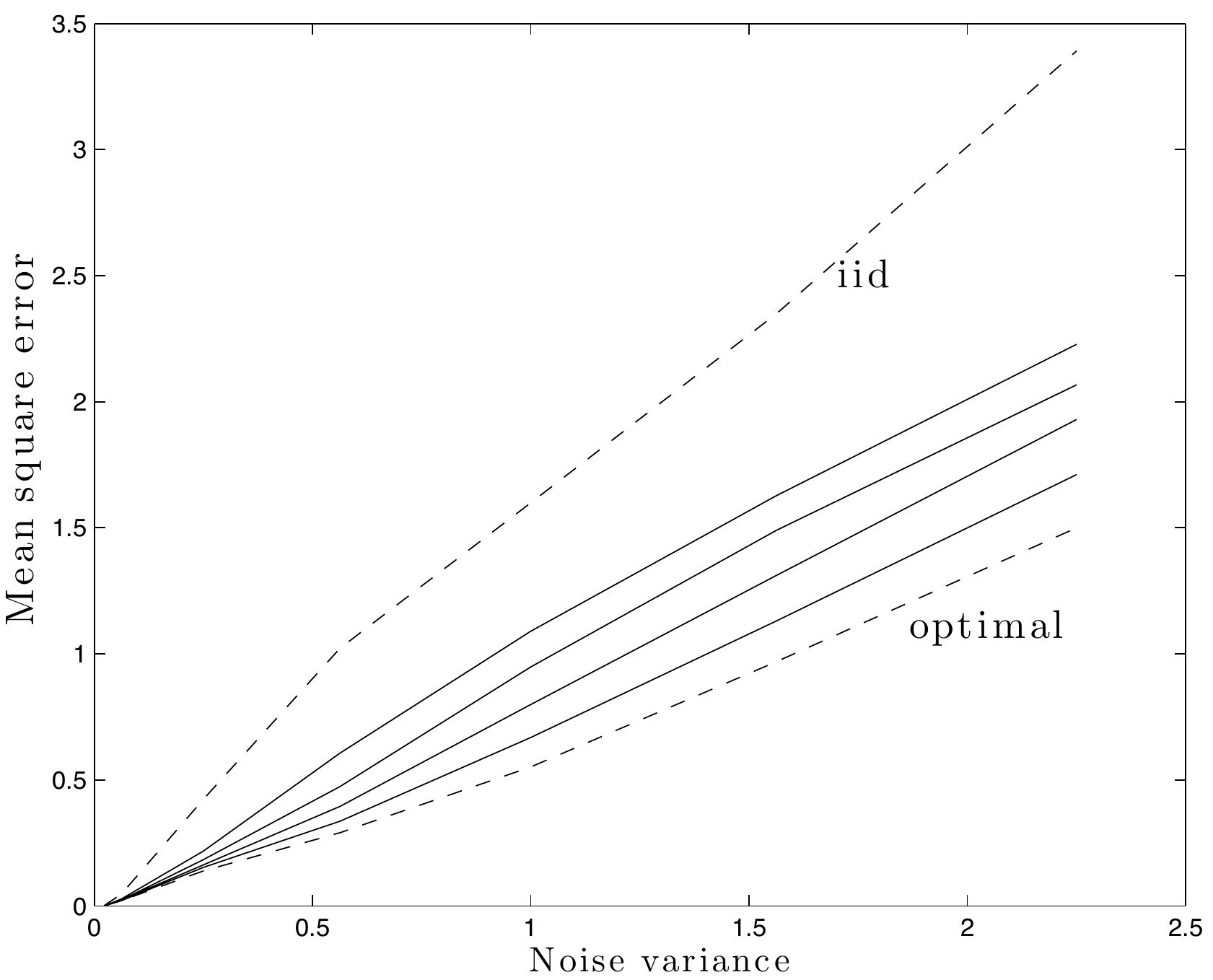}
\caption{Mean Square Error of lower bound reduced complexity filters computed using five different transition matrices $\ltp$ summarized in 
Table \ref{tab:ltp}. The transition matrix $\tp$ of dimension $3125 \times 3125 $ is specified in (\ref{eq:tp5}). The four solid lines (lowest to highest curve) are for  $\epsilon = 0.4,0.8,1.2,1.6$.
The optimal filter corresponds to $\epsilon = 0$, while the iid approximation corresponds to $\epsilon = 2$.}
\label{fig:mse}
\end{figure}

At each time $k$,
the reduced complexity filter $\lbelief_k = \filter(\lbelief_{k-1},\obs_k;\ltp)$ incurs computational cost of $O(\statedim \lowdim)$ where $\statedim = 3125$ and $\lowdim$ is specified
in Table \ref{tab:ltp}.
For each matrix $\ltp$ and noise variances $\onoisevar$ in the range $(0,2.25] $ we ran the reduced complexity HMM filter $\filter(\belief,\obs;\ltp)$ for  a million iterations and computed the average mean square error
of the state estimate. These average mean square error values are displayed in Fig.\ref{fig:mse}.
As 
might be intuitively expected,  Fig.\ref{fig:mse} shows that the reduced complexity filters yield a mean square error that lies between the iid approximation ($\epsilon = 2$)
and the optimal filter ($\epsilon = 0$). In all cases, as mentioned in Theorem \ref{thm:main}, the estimate $\lbelief_k $ provably lower bounds the true posterior $\belief_k$ as
$\lbelief_k \lr \belief_k$ for all time $k$.
Therefore the conditional mean estimates satisfy $\lmean_k \leq \mean_k$ for all $k$.

\begin{figure}
\centering
\subfigure[$\epsilon = 2$]{\includegraphics[width=0.45\columnwidth]{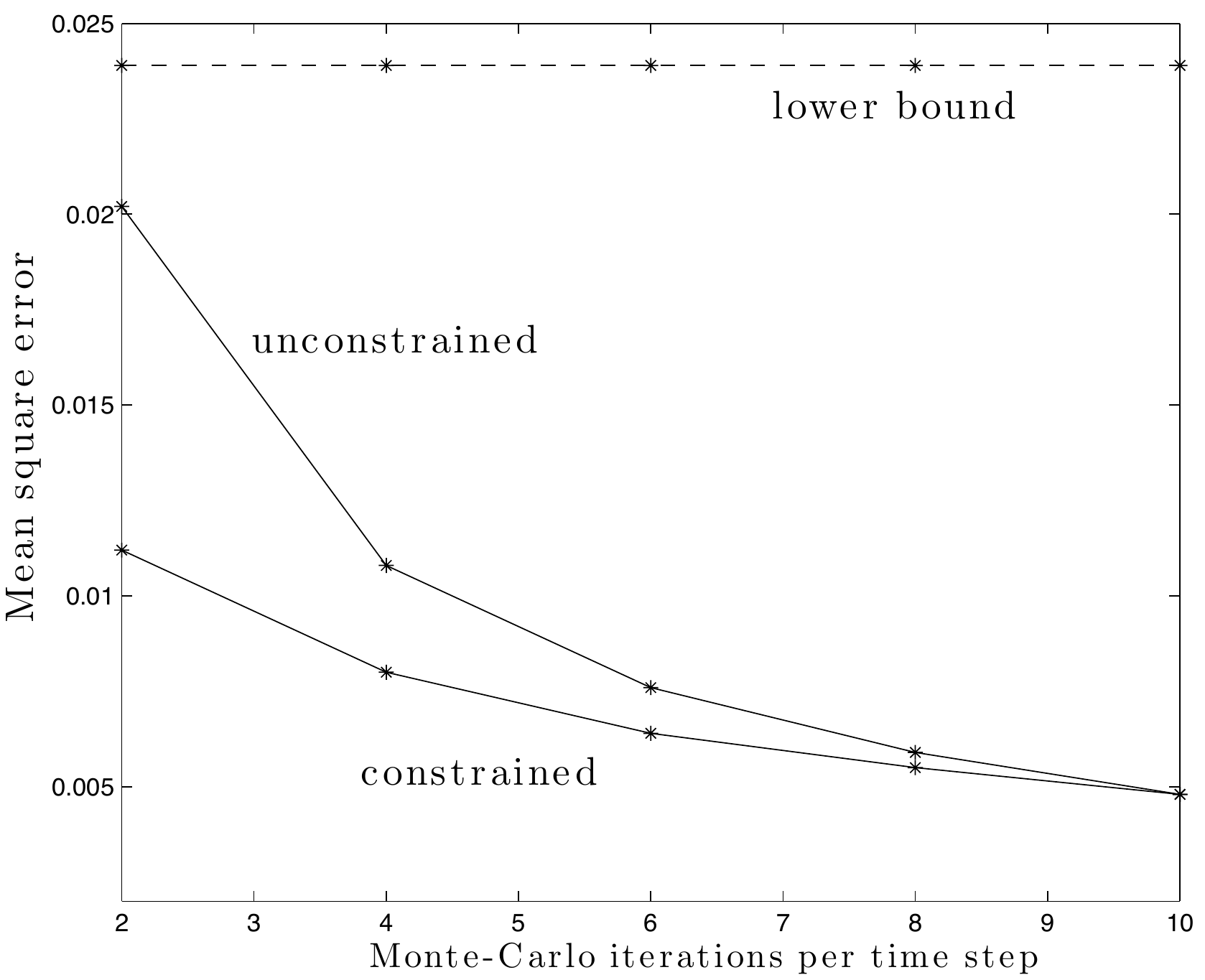}}
\subfigure[$\epsilon=1.6$]{\includegraphics[width=0.45\columnwidth]{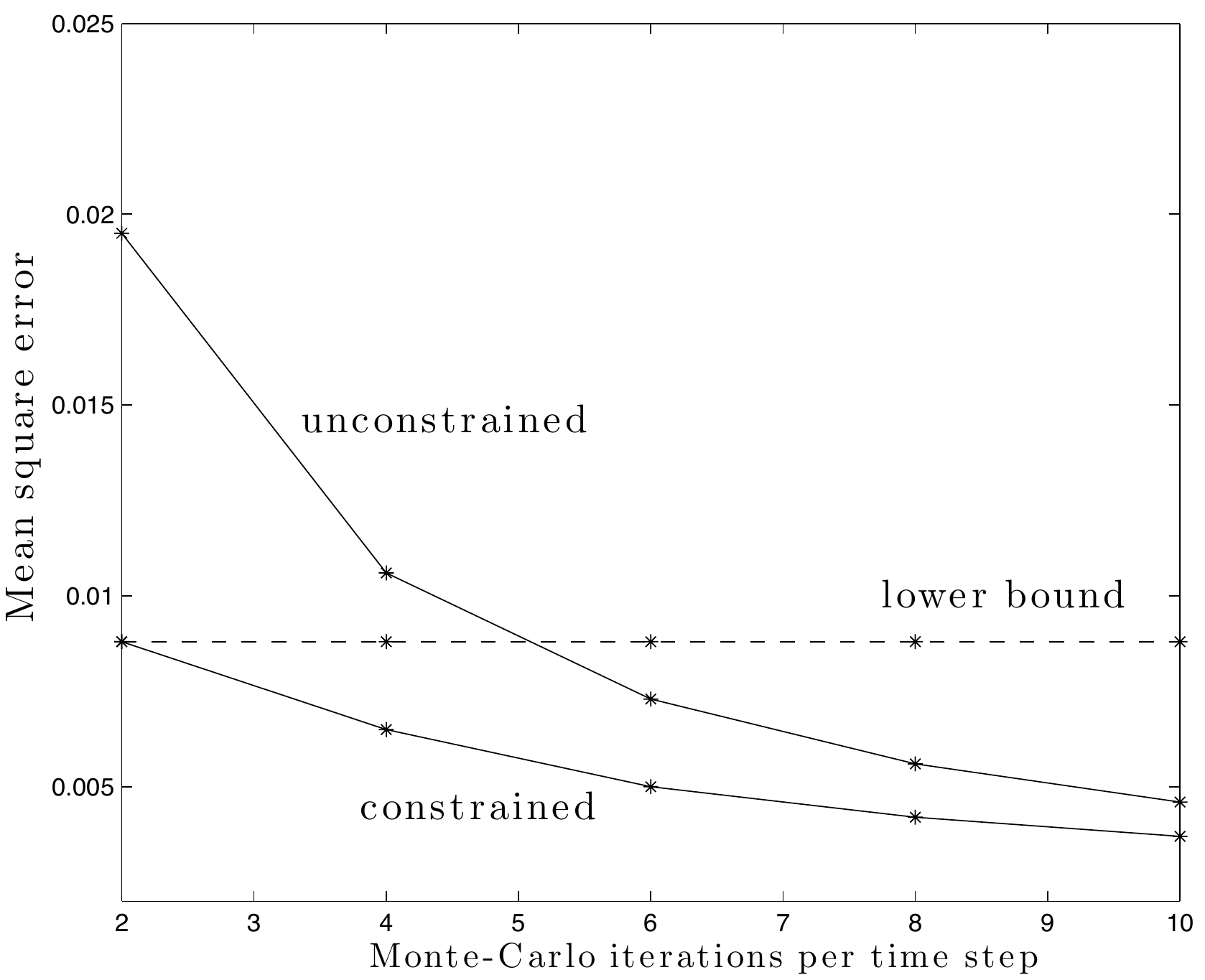}}

\caption{Mean Square Error between optimal predictor and stochastic dominance constrained importance sampling predictor of Algorithm \ref{alg:pf}.
Also shown is the mean square error of the unconstrained importance sampling predictor and the mean square error of the lower bound reduced complexity predictor.
The transition matrix $\tp$ is $3125 \times 3125 $ is specified in (\ref{eq:tp5}).
The reduced complexity predictor for  $\epsilon =2$ corresponds to the iid transition matrix $\ltp$ of rank 1, while
$\epsilon=1.6$ corresponds to $\ltp$ of rank 40; see Table \ref{tab:ltp}.}
\label{fig:mc}
\end{figure}

\subsubsection{Stochastic Dominance Constrained Importance Sampling  Algorithm  \ref{alg:pf}}
 Recall Algorithm~\ref{alg:pf}  computes the predicted posterior  $\hbelief_{k|k-1}$ by exploiting the lower and upper bound stochastic dominance constraints.
 To illustrate the performance of Algorithm \ref{alg:pf},  we 
 computed the mean square error between the estimated predictor using Algorithm \ref{alg:pf} and optimal predictor, that is,
 $(\hbelief_{k|k-1} - \belief_{k|k-1})^2 $ averaged over a million belief states $\belief_{k-1}$ sampled uniformly from the $5^5 -1$ dimensional  unit simplex.

We ran Algorithm~\ref{alg:pf}  for 5 different values of $L$, namely, 2,4,6,8,10 iterations at each time
step. Naturally, the more iterations $L$ per time step, the more accurate the estimate.
 Fig.\ref{fig:mc}(a) and \ref{fig:mc}(b) display these mean square errors for the constrained importance sampling filter for 5 values of $L$.  
 Fig.\ref{fig:mc}(a) corresponds to
  $\epsilon = 2$, resulting in $\ltp$ of rank 1.  Fig.\ref{fig:mc}(b) corresponds to $\epsilon = 1.6$, resulting in $\ltp$ of rank 40.
  Recall the performance of the lower bound estimates with these transition
  matrices were 
reported in Sec.\ref{sec:lcf}. 
  Fig.\ref{fig:mc}(a) and \ref{fig:mc}(b)   also display the mean square error  of the unconstrained
 importance sampling filtering algorithm which does not exploit the stochastic dominance constraints.  The dashed lines in the figures correspond
  to the mean square errors of the lower bound predictor $\lbelief_{k|k-1}$.
The figures show that  reductions in the mean square error occur by exploiting the stochastic dominance constraints; even for the  iid 
lower bound case ($\epsilon = 2$).

\subsubsection{Explicit Bounds} 
\begin{figure}
\centering
\includegraphics[width=0.7\columnwidth]{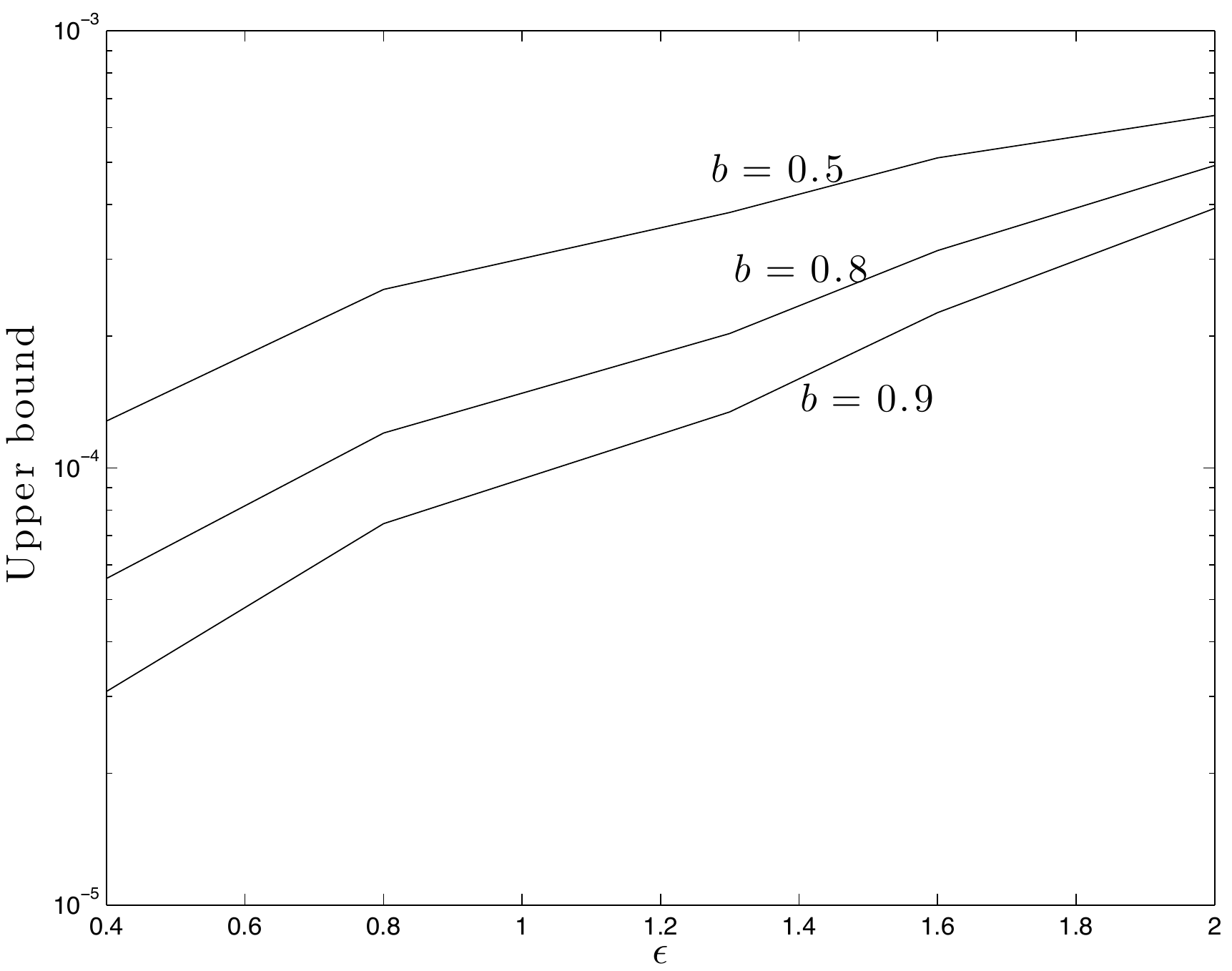}
\caption{The "upper bound" in the figure denotes the  right hand side of (\ref{eq:ebound})   
normalized by $\|g\|_1$. The values displayed are for 
five different values of $\epsilon$ corresponding to five different transition matrices $\ltp$  whose ranks are given  in 
Table \ref{tab:ltp}.
The observation matrix parametrized by $b$ is specified in (\ref{eq:oprobb}).}
\label{fig:dob}
\end{figure}

We now illustrate the explicit bound (\ref{eq:ebound}). We chose the same 3125 state Markov chain
with $\ltp, \tp$ as above and
a tridiagonal observation
matrix 
\beq  \oprob_{\state \obs} = \begin{cases}  b &  \text{ if } y = x  \\
\frac{1}{2} (1-b)  & \text{ if }  y = x-1  \text{ or } y = x+1. \end{cases} 
\label{eq:oprobb}
\eeq
We evaluated the  right hand side of the bound (\ref{eq:ebound}) normalized by $\|\levels\|_1$
for 5 different choices of $\epsilon\in \{0.4,0.8,1.2,1.6,2\}$ defined
in constraint (\ref{eq:con2}). (Recall from Table \ref{tab:ltp} that these correspond to 5 different choices of $\ltp$.)
Fig \ref{fig:dob} displays these bounds for three different observation matrices, namely $b = 0.9$, $b=0.8$ and $ b = 0.5$.
The figure shows that the bounds have two properties that are intuitive:
First  as  $\epsilon$ get smaller, the  approximation $(\ltp-\tp)^\p \pi$ gets tighter   and so one would expect that
$\E_{\obs}  \left| \levels^\p \left( \filter(\belief,\obs;\tp) - \filter(\belief,\obs;\ltp) \right) \right|$ is smaller. This is reflected
in  the upper bound displayed in the figure. 
Second,  for larger values of $b$, the "smaller" the noise and so the higher the estimation accuracy. Again the bounds reflect this.

\section{Discussion} The main idea of the  paper is to develop reduced complexity HMM filtering algorithms with provable sample path bounds. At each iteration,
the optimal HMM filter has $O(\statedim^2)$ computations and our aim was to derive reduced complexity upper and lower bounds with complexity $O(\statedim \lowdim)$ where $\lowdim \ll
\statedim$.
The paper is comprised of 4  main results. Theorem \ref{thm:main} showed that   one can construct transition matrices $\ltp$ and $\utp$
and lower and upper bound beliefs $\belief_k$ and $\ubelief_k$ that sandwich the true posterior $\belief_k$ as $\lbelief_k \lr \belief_k \lr \ubelief_k$,  for all time $ k = 1,2,\ldots $. Theorem~\ref{thm:tp2}  generalizes this to multivariate
TP2 orders. Sec.\ref{sec:convex} used copositive programming methods to construct low rank transition matrices $\ltp$ and $\utp$ of rank $\lowdim$ by minimizing the nuclear norm to guarantee 
$\|\ltp^\p \belief - \tp^\p \belief\|_1 \leq \epsilon$ and $\|\utp^\p \belief - \tp^\p \belief\|_1 \leq \epsilon$ over the space of all posteriors $\Belief$. Finally, Theorem \ref{thm:copositive} derived
explicit bounds between the optimal estimates and the reduced complexity estimates. 

It is interesting that the derivation of MLR stochastic dominance bounds in this paper involves  copositivity conditions.
There is a rich literature in copositivity including computational aspects \cite{BD08,BD09}. In future work it is worthwhile  extending the bounds in this paper to copositive kernels
for continuous state filtering problems. Such results could yield guaranteed sample path bounds for general nonlinear filtering problems.

\appendix

\section{Proofs}

\subsection{Proof of Theorem \ref{thm:main}}
1. By definition, $\tp$ being TP2 implies its rows $\tp_i$ satisfy, $\tp_1 \lr  \tp_2 \cdots \lr \tp_\statedim$.
Choose $\ltp$ such that its rows satisfy $\ltp_i  \lr \tp_i$ for all $i=1,2,\ldots,X$. Then it is straightforward to show that $\ltp \lR \tp$.
Similarly choosing the rows of $\utp$ as $\utp_i \gr \tp_\statedim$ for $i=1,2,\ldots,X$ implies that $\utp \gR \tp$.

2. By definition
${\tp}^\p \pi \gr {\ltp}^\p \pi$ is equivalent to
$$\sum_{i} \sum_{m} \left(\tp_{ij} \ltp_{m,j+1} - \ltp_{ij}\tp_{m,j+1}\right) \pi_i \pi_m \leq 0 $$
for $j=1,\ldots,\statedim$. Finally, it is straightforwardly verified that 
$\belief \gr \lbelief$ implies $\frac{\oprob_{\obs} \belief}{\one^\p \oprob_\obs \belief}  \gr  \frac{\oprob_{\obs} \lbelief} {\one^\p \oprob_\obs \lbelief}$.
(In fact it is this crucial property of closure under Bayes' rule that makes the MLR stochastic order ideal for the results in this paper).

3. Suppose $\lbelief_k \lr \belief_k$. Then by Statement 2,
$\filter(\lbelief_k, \obs_{k+1}; \ltp) \lr \filter(\lbelief_k, \obs_{k+1}; \tp)$. 
Next since $\tp$ is TP2,  it follows that $ \lbelief_k \lr \belief_k$ implies  $\filter(\lbelief_k, \obs_{k+1}; \tp) \lr \filter(\belief_k, \obs_{k+1}; \tp)$.
Combining the two inequalities yields
$\filter(\lbelief_k, \obs_{k+1}; \ltp) \lr  \filter(\belief_k, \obs_{k+1}; \tp)$, or equivalently $\ltp_{k+1} \lr \tp_{k+1}$.
Finally, MLR dominance implies first order dominance which by Result \ref{res1} implies  dominance of means thereby proving 3(a).

To prove 3(b) we need to show that 
$\lbelief \lr \belief $ implies $\arg\max_ i \lbelief(i) \leq \arg\max_i\belief(i)$.  This is shown by contradiction:
Let $i^* = \argmax_i \belief_i$ and $j^* = \argmax_j \lbelief_j$.  Suppose $i^* \leq j^*$. Then $\belief \gr \lbelief $
implies $\belief(i^*) \leq \frac{\lbelief(i^*)}{\lbelief(j^*)} \belief(j^*) $. Since $\frac{\lbelief(i^*)}{\lbelief(j^*)} \leq 1$, we have 
 $\belief(i^*) \leq  \belief(j^*) $ which is a contradiction since $i^*$ is the argmax for $\belief(i)$.

\subsection{Proof of Theorem \ref{thm:tp2}}
If suffices to show that $\uA \lR A \implies \uA \otimes \uA \lR A  \otimes A$. (The proof for repeated Kronecker products then follows straightforwardly by induction.)
Consider the TP2 ordering in Definition~\ref{def:tp2o}.
The indices $\i=(j,n)$ and $\j = (f,g)$ are each two dimensional. 
There are four cases: $(j<f, n <g)$, $(j<f, n>g)$, $(j>f, n<g)$, $(j>f, n>g)$.
TP2 dominance for the first and last cases are trivial to establish. We now show  TP2 dominance for the third case (the second case follows similarly):
Choosing the indices $\i = (j, g-1)$ and $\j = (j-1,g)$,
 it follows that  $\uA \otimes \uA \lR A  \otimes A$ is equivalent to
$$ \sum_m \sum_l A_{m,g-1} \uA_{l,g} \sum_i \sum_k  (A_{ij} \uA_{k,j+1} - A_{i,j+1}\uA_{k,j})  \belief_{im} \belief_{kl} \leq 0$$
So a sufficient condition is that for any non-negative numbers $\belief_{im}  $ and $\belief_{kl}$, 
$\sum_i \sum_k  (A_{ij} \uA_{k,j+1} - A_{i,j+1}\uA_{k,j})  \belief_{im} \belief_{kl}  \leq 0$ which is equivalent to $\uA \lR A$ by Definition \ref{def:lR}.

\subsection{Proof of Theorem \ref{thm:copositive}}

We start with the following theorem that characterizes the $l_1$ (equivalently, variational distance) in the classical Bayes' rule. Recall that the Bayes' rule
update using prior $\belief$ and observation $\obs$ is
$$ \bayes(\belief,\obs) = \frac{\oprob_\obs \belief}{\one^\p \oprob_\obs \belief} .$$
(Of course this is the same as the optimal filter with transition operator being identity).

 \begin{theorem}  \label{thm:bayesbackvar}
 Consider any two posterior  probability mass functions $\belief, \tbelief \in \Belief$.
Then: 
\begin{compactenum}
\item  The variational distance in the Bayesian update satisfies
  $$ \dvar{ \bayes(\belief,\obs) } { \bayes(\tbelief,\obs) } \leq \frac{\max_{i} \oprob_{i,\obs}} {\one^\p \oprob_\obs \belief  }  \dvar{\belief}{\tbelief}. $$
  (Recall that the variational distance is half the $l_1$ norm).
\item  The normalization term in Bayes' rule satisfies
$$\one^\p  \oprob_\obs \belief  \geq  \max\{ \one^\p  \oprob_\obs \tbelief - \epsilon  \max_i \oprob_{iy},  \min_i \oprob_{iy} \}.$$
\end{compactenum}
 \end{theorem}

{\bf Proof}: We refer to \cite{CMR05} for a textbook treatment of similar proofs on more general spaces.
\subsubsection{Statement 1}
 For any $g\in \reals^\statedim$,
\begin{align}
& g^\p\(\bayes(\belief, \obs) - \bayes(\tbelief, \obs) \) \nn \\ &= g^\p \(\bayes(\belief, \obs) - \frac{\oprob_\obs  \tbelief}{ \one^\p \oprob_\obs \belief} +   \frac{\oprob_\obs  \tbelief}{ \one^\p \oprob_\obs \belief} - \bayes(\tbelief, \obs)\) \nonumber \\
&= \frac{1}{\one^\p \oprob_\obs \belief} \, g^\p \left[ \identity - \bayes(\tbelief, \obs) \one^\p \right] \oprob_\obs\, (\belief - \tbelief). 
\label{eq:dobstep}
\end{align}
Applying the result\footnote{This inequality is tighter than Holder's inequality which is
$|f^\p (\belief - \lbelief)| \leq 2 \max_i |f_i|\, \dvar{\belief}{\tbelief}$.}
 that for any vector $f \in \reals^\statedim$, 
 \beq | f^\p (\belief - \lbelief) | \leq \max_{i,j} |f_i - f_j| \dvar{\belief}{\lbelief} \label{eq:tightholder}\eeq
  to the right hand side of the above equation yields, 
$$ | g^\p \bigl(\bayes(\belief, \obs) - \bayes(\tbelief, \obs)\bigr) | \leq \frac{1}{\one^\p \oprob_\obs \belief} \, \max_{i,j} |f_i - f_j| \, \dvar{\belief}{\tbelief} $$
where $f_i = g^\p  \left[ \identity - \bayes(\tbelief, \obs) \one^\p \right] \oprob_\obs e_i$ and  $f_j = g^\p  \left[ \identity - \bayes(\tbelief, \obs) \one^\p \right] \oprob_\obs e_j$.

So $$|f_i - f_j| = |g_i \oprob_{i,\obs}  - g^\p \bayes(\tbelief, \obs) \oprob_{i,\obs} -  ( g_j \oprob_{j,\obs} - g^\p \bayes(\tbelief, \obs) \oprob_{j,\obs} ) |.$$
Since  $\bayes(\tbelief,\obs)$ is a probability vector, clearly $|g^\p  \bayes(\tbelief, \obs) | \leq  \max_i |g_i|$. This  together with the fact that $\oprob_{i,\obs}$ are non-negative implies
$$\max_{i,j} | f_i - f_j | \leq 2 \max_i |g_i| \max_i  \oprob_{i,\obs}. $$
So denoting $\|g\|_\infty = \max_i|g_i|$,  we have
$$ |g^\p\(\bayes(\belief;\oprob_\obs) - \bayes(\tbelief;\oprob_\obs) \) |  \leq  2 \, \frac{\|g\|_\infty \max_{i} \oprob_{i,\obs}} {\one^\p \oprob_\obs \belief  }  \dvar{\belief}{\tbelief} .$$
Finally applying 
the result that $\|f \|_1  = \max_{\|g\|_\infty = 1} |g^\p f| $ for $g\in \reals^\statedim$ (see  \cite[pp.267]{HJ12}), yields
\begin{align*}
  \|\bayes(\belief, \obs) - \bayes(\tbelief, \obs)\|_1 & = \max_{\|g\|_\infty = 1} |g^\p\(\bayes(\belief, \obs) - \bayes(\tbelief, \obs) \)|  \\
  &\leq  \max_{\|g\|_\infty = 1}     2 \, \frac{\|g\|_\infty \max_{i} \oprob_{i,\obs}} {\one^\p \oprob_\obs \belief  }  \dvar{\belief}{\tbelief}.
  \end{align*}

 \subsubsection{Statement 2} 
 
Applying Holder's inequality yields
$$ |\one^\p \oprob_y (\belief - \tbelief) | \leq \|\one^\p \oprob_y \|_\infty  \|\belief - \tbelief\|_1 = \max_i \oprob_{iy} \, \epsilon$$
 implying that 
 \beq  \one^\p \oprob_y \belief \geq  \one^\p \oprob_y \tbelief  - \epsilon  \max_i \oprob_{iy}.  \label{eq:holder} \eeq
 Also clearly $ \one^\p  \oprob_{\obs}\belief \geq  \min_i \oprob_{iy} \one^\p  \belief  = \min_i \oprob_{iy}$. Combining this with (\ref{eq:holder}) proves the result.

\subsubsection{Proof of Theorem \ref{thm:copositive}} With the above results we are now ready to prove the theorem. The triangle inequality for norms  yields
\begin{align}
& \dvar{\belief_{k+1} } { \lbelief_{k+1} }   = \dvar{ \filter(\belief_k,\obs_{k+1};\tp) }{ \filter(\lbelief_k,\obs_{k+1};\ltp) } \nonumber  \\
& \leq  
\dvar{ \filter(\belief_k,\obs_{k+1};\tp) } { \filter(\belief_k,\obs_{k+1};\ltp) } \nn \\ &  \hspace{1cm} + \dvar{ \filter(\belief_k,\obs_{k+1};\ltp) } {  \filter( \lbelief_k,\obs_{k+1}; \ltp)}. 
\label{eq:triangle}
\end{align}

{\bf Part 1}: Consider the first normed term in the right hand side of (\ref{eq:triangle}).
Applying (\ref{eq:dobstep}) with 
  $\belief = \tp^\p \belief_k$ and $\tbelief = \ltp^\p \belief_k$ yields
\begin{multline*}
  \levels^\p (\filter(\belief_k,\obs;\tp) - \filter(\belief_k,\obs;\ltp)) \\ = \frac{1}{\filterd(\belief,\obs;\tp)}
  \levels^\p \left[ I - \filter(\belief,\obs,\ltp) \one^\p\right] \oprob_\obs (\tp - \ltp)^\p \belief  
\end{multline*}
where $\filterd(\belief,\obs;\tp) = \one^\p  \oprob_\obs \tp^\p \belief$.
  Then (\ref{eq:tightholder}) yields 
\begin{multline*}  \levels^\p (\filter(\belief_k,\obs;\tp) - \filter(\belief_k,\obs;\ltp))\\  \leq 
\max_{i,j}  \frac{1}{\filterd(\belief,\obs;\tp)}  \levels^\p \left[ I - \filter(\belief,\obs,\ltp) \one^\p\right] \oprob_\obs (e_i - e_j) \dvar{\tp^\p \belief}{ \ltp^\p \belief} \end{multline*}
Since $\dvar{\tp^\p \belief}{ \ltp^\p \belief} \leq \epsilon$, taking expectations with respect to the measure $\filterd(\belief,\obs;\tp)$, completes the proof
of the first assertion.
  
{\bf   Part 2}:
Applying Theorem \ref{thm:bayesbackvar}(i) with the notation   $\belief = \tp^\p \belief_k$ and $\tbelief = \ltp^\p \belief_k$ yields
\begin{align}  &\dvar{\filter(\belief_k,\obs;\tp) } { \filter(\belief_k,\obs;\ltp) } \leq 
 \frac{ \max_i \oprob_{i,\obs}   \dvar{\tp^\p \belief_k}{\ltp^\p \belief_k} } { \one^\p  \oprob_{\obs} \ltp^\p \belief_k }  \nonumber\\
& \leq  \frac{\epsilon}{2}\,\frac{ \max_i \oprob_{i,\obs} } { \one^\p  \oprob_{\obs} \ltp^\p \belief_k } 
\leq   \frac{ \max_i \oprob_{i,\obs}  \, \epsilon/2} { \max\{ \one^\p  \oprob_{\obs} \ltp^\p \lbelief_k - \epsilon \max_i \oprob_{i\obs}, \min_i \oprob_{iy}  \}}.
\label{eq:term1}
\end{align}
The second  last inequality follows from the construction of $\ltp$ satisfying  (\ref{eq:con2})  (recall the variational norm is half the $l_1$ norm). The last inequality follows from 
Theorem \ref{thm:bayesbackvar}(ii).

Consider the second normed term in the right hand side of (\ref{eq:triangle}).
Applying Theorem  \ref{thm:bayesbackvar}(i) with notation $\belief = \ltp^\p \belief_k$ and $\tbelief = \ltp^\p \lbelief_k$ yields
\begin{multline} \dvar{ \filter(\belief_k,\obs;\ltp) } {  \filter( \lbelief_k,\obs; \ltp) } 
 \leq 
 \frac{ \max_i \oprob_{i,\obs}   \dvar{\ltp^\p \belief_k}{\ltp^\p \lbelief_k} } { \one^\p  \oprob_{\obs} \ltp^\p \lbelief_k } \\ \leq
\frac{ \max_i \oprob_{i,\obs} \, \dob(\ltp) \, \dvar{ \belief_k}{\lbelief_k} } { \one^\p  \oprob_{\obs} \ltp^\p \lbelief_k } \label{eq:term2}
  \end{multline}
 where the last inequality follows from the submultiplicative property of the Dobrushin coefficient.
Substituting   (\ref{eq:term1}) and  (\ref{eq:term2}) into the right hand side of the triangle inequality (\ref{eq:triangle}) proves the result.

\bibliographystyle{IEEEtran}
\bibliography{$HOME/styles/bib/vkm}
\end{document}